\documentclass[12pt, preprint]{aastex}

\begin{document}

\def\d{{\rm d}}
\title{Evolution of the Loop-Top Source of Solar Flares---Heating and Cooling Processes}

\author{Yan Wei Jiang\altaffilmark{1}, Siming Liu\altaffilmark{1}, Wei Liu\altaffilmark{1} and 
Vah\'{e} Petrosian\altaffilmark{2}}

\affil{Center for Space Science and Astrophysics, Stanford University, Stanford, California
94305}

\altaffiltext{1}{Department of Physics; arjiang@stanford.edu, liusm@stanford.edu, 
weiliu@sun.stanford.edu, }
\altaffiltext{2}{Departments of Physics and Applied Physics; vahe@astronomy.stanford.edu}

\begin{abstract}

We present a study of the spatial and spectral evolution of the loop-top (LT) sources in a sample 
of 6 flares near the solar limb observed by {\it RHESSI}.  A distinct coronal source, which we 
identify as the LT source, was seen in each of these flares from the early ``pre-heating'' phase 
through the late decay phase. Spectral analyses reveal an evident steep power-law component in the 
pre-heating and impulsive phases, suggesting that the particle acceleration starts upon the onset 
of the flares.  In the late decay phase the LT source has a thermal spectrum and appears to be 
confined within a small region near the top of the flare loop, and does not spread throughout the 
loop, as is observed at lower energies. The total energy of this source decreases usually faster 
than expected from the radiative cooling but much slower than that due to the classical Spitzer 
conductive cooling along the flare loop. These results indicate the presence of a distinct LT 
region, where the thermal conductivity is suppressed significantly and/or there is a continuous 
energy input. We suggest that plasma wave turbulence could play important roles in both heating 
the plasma and suppressing the conduction during the decay phase of solar flares. With a simple 
quasi-steady loop model we show that the energy input in the gradual phase can be 
comparable to that in the impulsive phase and demonstrate how the observed cooling and confinement 
of the LT source can be used to constrain the wave-particle interaction.

\end{abstract}


\section{INTRODUCTION}

The properties of thermal and nonthermal emissions of solar flares and their correlation are 
very important in understanding the flare energizing process. The well-known
Neupert effect (Neupert 1968) states that the time integral of the hard X-ray (HXR) or
microwave emission is correlated with the light curve of the soft X-ray (SXR) emission.  
This effect is generally explained with the non-thermal thick-target model, where flares
are triggered by injections of nonthermal electrons that produce the HXRs and deposit most of 
their energy at the footpoints (FPs). This causes chromospheric evaporation filling the loop 
with hot thermal plasmas responsible for the observed SXRs (Brown 1971; Lin \& Hudson 
1971; Petrosian 1973;  McTiernan et al. 1999). 
However, detailed studies (Dennis \& Zarro 1993; Veronig et al. 2002) show deviations from this
simple effect, indicating that this picture is not complete. There is clear evidence of a SXR
rise before the impulsive HXR emission of some flares (often referred to as the pre-heating), 
and a persistence of the increase or slower than expected decrease of the SXR flux is 
frequently observed after the HXR emission has died 
out.  Statistical studies (e.g., Lee, Petrosian \& McTiernan 1995) indicate a heating process other 
than the collisional heating mentioned above. 

Since the discovery of impulsive HXR emission from the loop-top (LT) by {\it Yohkoh} (Masuda et
al. 1994; Masuda 1994), it has been realized that the spatial and spectral evolution of the LT
source plays a crucial role in revealing the energy release process of magnetic reconnection and
the consequent particle acceleration and plasma heating in solar flares (Tsuneta 1996; Alexander
\& Metcalf 1997; Petrosian, Donaghy \& McTiernan 2002). To explain these observations one needs
to trap high-energy electrons in the LT region. However, due to the low energy resolution of the
{\it Yohkoh} HXR Telescope (HXT), the exact nature of the LT and FP sources is not
well-constrained. It is, e.g., difficult to tell whether the LT source has a steep nonthermal 
spectrum or is produced by a ``superhot'' thermal gas (Lin \& Schwartz 1987; Tsuneta et al. 1997; 
Petrosian et al. 2002). Analyses of unresolved observations during the
late decay phase have shown that the observed energy decline does not agree with that expected
through conduction (Moore et al 1980; McTiernan et al. 1993). Several mechanisms may alleviate
this discrepancy (Antiochos \& Sturrock 1976, 1978; Rosner et al. 1986; Takahashi \& Watanabe
2000; Reeves \& Warren 2002), however, these models were designed to explain observations with
low spectral, spatial and temporal resolutions. {\it RHESSI} with its higher resolutions can be
very helpful in this regard.

In this paper we present a systematic imaging and spectroscopy investigation of the LT source of 6 
simple {\it RHESSI} limb flares, each of which seems to be associated with a single loop, instead 
of multiple loops for complex flares.  In general, the spectrum of the LT source has two 
components, a low-energy thermal component plus a power-law tail (or a broken power-law). The 
nonthermal component dominates in the early phases. The thermal component becomes more and more 
prominent as the flare proceeds, and in the late decay phase the nonthermal component vanishes 
completely. From spectral fits we determine the evolution of the emission measure $EM$ and 
temperature $T$ of the LT source during the decay phase. We also determine the evolution of the 
volume of the source $V$. These give us the evolution of the density $n$ and the total energy 
${\cal E}=3nk_{\rm B} TV$, where $k_{\rm B}$ is the Boltzmann constant. From these we find that 
the energy decay rate is higher than the radiative cooling rate, but much lower than that due to 
the thermal conduction, consistent with previous studies (e.g., McTiernan et al.  1993). Clearly 
suppression of conduction and/or continuous energy release are required to explain these 
observations. Another important and surprising feature of these sources is that they are often 
found localized near the LT, which is distinct from observations at lower energies where emission 
along the whole length of the loop is seen in an arcade structure. We explore the possible 
mechanisms for the confinement of these sources and for the suppression of conductivity and show 
that the former alone rules out quasi-steady loop models with a uniform conductivity even when the 
radiative cooling is taken into account. We suggest that plasma waves or turbulence, which we 
believe plays a major role in the acceleration of particles during the impulsive phase (see e.g., 
Petrosian et al. 2002), can be the agent of both the heating and confining of the LT source during 
the decay phase as well. We consider a steady-state isobaric loop model with turbulence plasma 
waves concentrated in a distinct LT region and explore the degree to which the observed LT 
confinement can set constraints on the plasma heating and conduction suppression processes. 
The model predicts strong line emissions from the FPs in the gradual phase, and for flares with 
a relatively longer decay time the energy input in the decay phase may exceed that of the 
impulsive phase. This is in direct contradiction with the Neupert effect, which claims that the 
energy release is limited to the impulsive phase. Broadband observations, especially those in the 
optical and UV bands, should be able to test the model.


In \S\ \ref{obs} we present the analyses of these flares and the key observational features.  
Theoretical investigations of these results and their implications are discussed in \S\ 
\ref{ana}, where a viable model is also presented. \S\ \ref{con} summarizes the main 
conclusions of this study.

\section{OBSERVATION AND DATA ANALYSIS}
\label{obs}

In this section we first discuss the flare selection criteria. The spatial and spectral 
properties of six selected flares are then studied in detail.

\subsection{Flare Selection}
\label{data}

Because we are interested in the evolution of flares, a good time-coverage of the whole event 
is required. To reduce the instrumental effects, only events with their shutter state 
remaining unchanged are considered.  We also focus on relatively large flares with peak count 
rate $>300$ s$^{-1}$ per detector in the $6-12$ keV energy band to ensure a high image quality 
and reliable spectral results.  To study the spatial evolution of the LT source, the flares 
need to be close to the solar limb and have a relatively simple morphology.  Furthermore, to avoid 
spectral contaminations from the FP sources, flares with their LT source staying above the solar limb 
and their FPs completely occulted by the solar disk would be ideal candidates. We choose 
flares with heliocentric longitude larger than $75^\circ$.

We searched events observed by {\it RHESSI} in 2002 and found a total of six flares appropriate 
for this study. Two of them are limb flares with both their LT and two FP sources clearly seen, 
one is a partially occulted flare, where the LT source (above the limb) and one FP source were 
observed, and the three remaining flares are coronal events above the solar limb (see Figures 
\ref{images1.ps}-\ref{images3.ps}). For each flare we also examined the simultaneous {\it GOES} 
observations extending to lower energies, and {\it SOHO} EUV Imaging Telescope (EIT) 195 \AA\ 
observations to determine the loop geometry. From these and observations described below, we found 
that all of these flares appear to be associated with a single-loop structure.

\subsection{Images}
\label{img}

We obtained images of the flares with the PIXON algorithm, which has the best spatial resolution.  
Because the energy resolution of detector 2 is severely degraded, which introduces significant 
uncertainties to the source structure at a given energy range, only the front segment of detectors 
3-8 were used, achieving a resolution of $\sim 5\farcs0$ (Aschwanden et al. 2003).  These images 
reveal a distinct LT source present from the beginning of the flares through the late decay phase, 
whereas the FP sources, whenever observed, are seen only during the impulsive phase at higher 
energy HXRs.  Figures \ref{images1.ps}-\ref{images3.ps} show the 6-12 keV images of the six 
flares at the HXR peak (thin contours for the first three and grey scale for the rest) and a late period 
in the decay phase (thick contours). The peak time FP sources (at 51-57 keV, 34-39 keV, and 60-100 
keV, respectively) are indicated by the gray scale for the two limb flares and the partially 
occulted April 4 flare\footnote{Note that one of the high energy sources for this flare, which 
would be normally associated with the FPs, is above the solar limb.}. The plus signs indicate the 
emission centroid of the LTs. There are two flares on April 4, one of which is partially 
occulted labeled ``a'' and the other is a corona event labeled ``b'' (Figure \ref{images2.ps}). 
For the April 30 flare, there are two corona sources during the 
impulsive phase. The lower one is associated with the LT, while the higher one may indicate the 
other end of a reconnection current sheet (Sui \& Holman 2003). The dotted contour at the 
upper-right corner gives the FWHM of the PSF of each flares at 6-12 keV. We see that all of the LT 
sources are partially resolved.

Although the LT sources appear to shift systematically upward as the flares proceed, the motions 
are not significant. And the LT sources seem to be confined within a small region with no 
significant changes in the source morphology throughout the flare. These suggest that the flares 
may be associated with a single-loop. More importantly, the confinement of the image indicates 
that they are not in line with the source evolution expected in the thick-target chromospheric 
evaporation model (Mariska et al. 1989). On the other hand, all the flares studied here underwent 
a prominent preheating phase. It is possible that this preheating process raised the gas pressure 
in the flare loop and suppressed the post impulsive phase chromospheric evaporation effects 
(Emslie et al. 1992). The results are also distinct from flare images obtained at lower energies, 
such as those taken with the EIT and {\it Yohkoh} SXT ($1-2$ keV), where the complete arcade loop 
structure is often seen.

\begin{figure}[thb]
\epsscale{1}
\centerline{
\plottwo{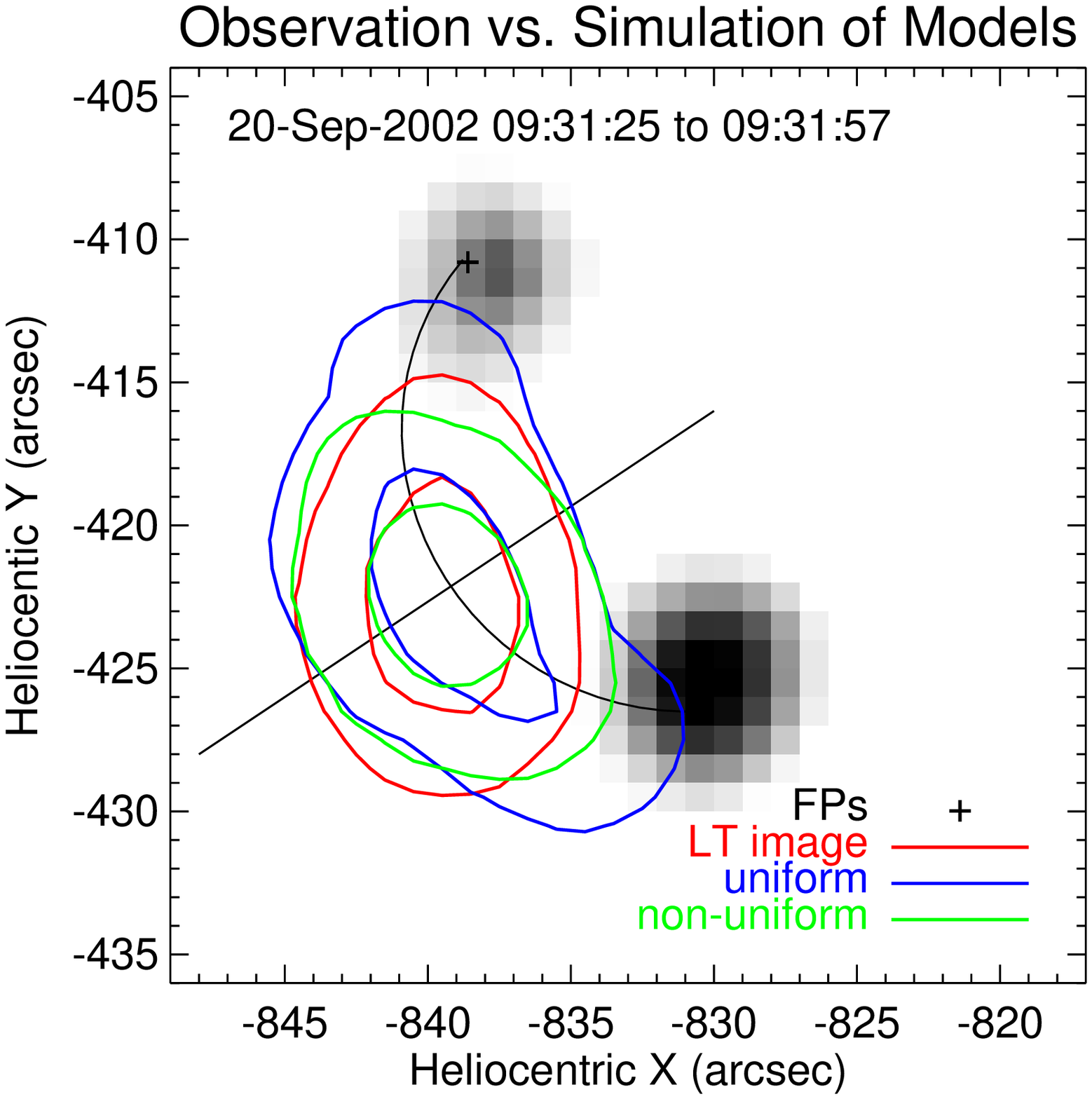}{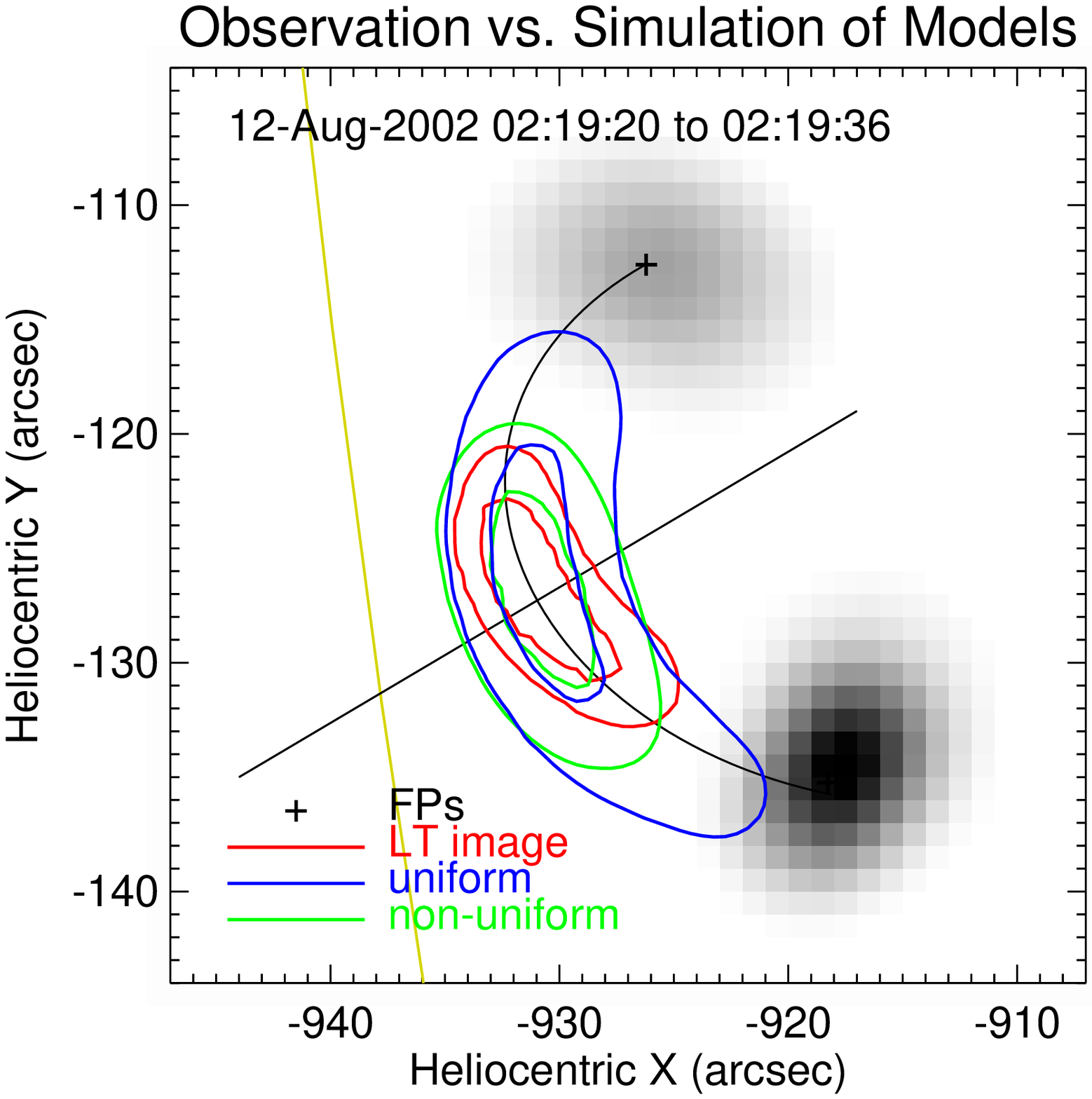}
}
\caption{
{\it Left:} Comparison of the simulated images with the observed image of the September 20 2002 
flare obtained with the PIXON algorithm. The gray scale shows the FPs (51-57 keV) at the HXR peak, and the 
red contours (15\% and 75\% at 6-12 keV) are for the LT with the time interval indicated.  The thin 
black lines indicate the loop 
structure (curved) and the direction in perpendicular to it (straight). The simulated image 
contours are for the uniform loop model (blue contours), which is longer than the observed 
images, and for the model with a distinct LT region (green contours), which 
agree with the observations. {\it Right:} Same as the {\it Left} but for the August 12 2002 flare.
} 
\label{model1.ps}
\end{figure}

\begin{figure}[thb]
\epsscale{1}
\centerline{
\plottwo
{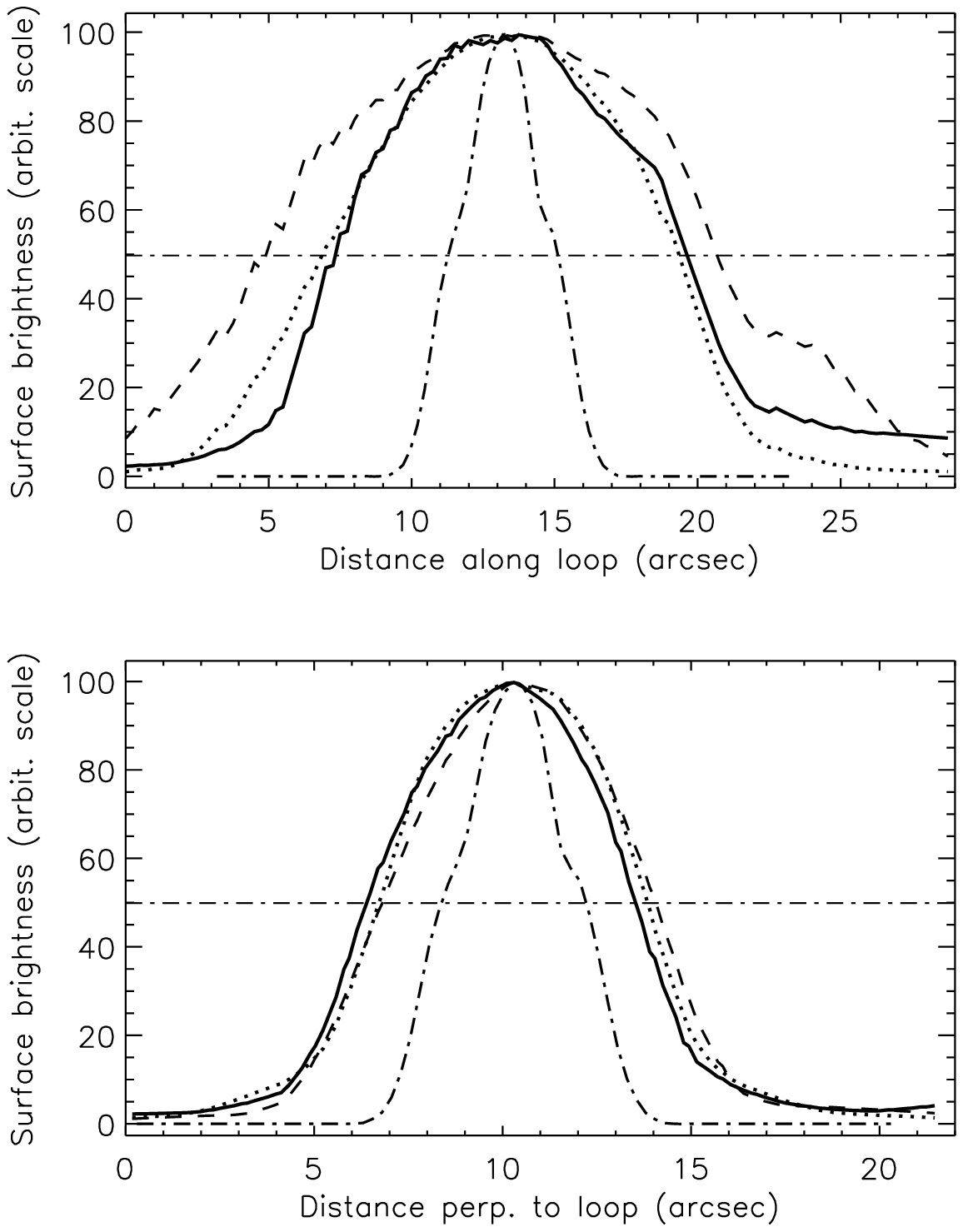}
{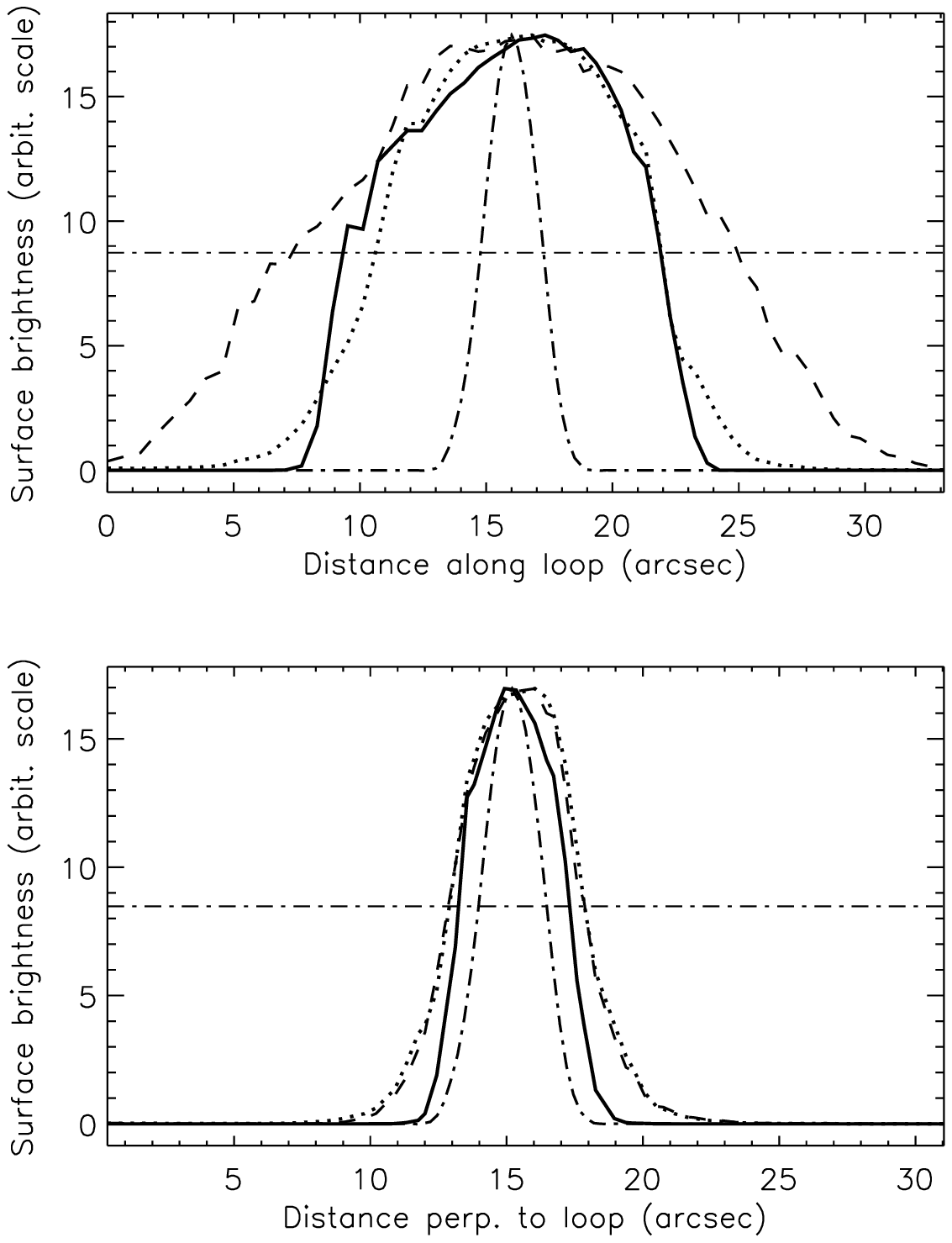}
}
\caption{
{\it Left:} Comparison of the observed and simulated image brightness profiles along (top) and in 
perpendicular to (bottom) the flare loop. The solid, dotted, dashed, and dot-dashed lines are for 
the observation, the non-uniform, uniform loop models, and the PSF, respectively. The LT is 
partially resolved in both directions. {\it Right:} Same as the {\it Left} but for the August 12 
2002 flare. The LT is resolved along the loop, but the profile in perpendicular to the loop is 
consistent with the PSF.
}
\label{model2.ps}
\end{figure}

For quantitative analyses we assume that the LT source is an ellipsoid and determine its 
semi-minor (perpendicular to the loop) and semi-major (along the loop) axes $a$ and $b$ by 
fitting the $15\%$ contour of the observed peak emission (outer contours in Figures 
\ref{images1.ps}-\ref{images3.ps} and red outer contours in Figure \ref{model1.ps}). To compare 
the observed source image with the theoretical loop models discussed below, we assume that the 
loop is a half circle and slightly tilted with respect to the solar surface so that the apex of 
the loop falls at the observed LT position, and the two FPs are at the observed FP emission 
centroids. The black lines in Figure \ref{model1.ps} show this loop structure (curved) and 
the direction perpendicular to it (straight).  In Figure \ref{model2.ps} the brightness profiles
along (upper) and in perpendicular to (lower) the loop are shown by the solid lines, the 
dot-dashed lines give the PSF. We see that the September 20 flare is resolved. The August 12 flare is 
resolved along the loop, but not resolved in the perpendicular direction. Because of this the 
size $b$ and specially $a$ should be considered as upper limits so that the volumes (densities) 
deduced below are upper (lower) limits. 

\subsection{Total Spectrum and Imaging Spectroscopy}
\label{spec}

For flares with their FPs completely occulted by the solar disk, the spectral evolution of the LT 
source over the entire flare duration can be readily studied with the spectral software package 
SPEX. The dominance of the LT source during the rising and decay phases also facilitates studying 
the spectrum of the LT source without employing the complicated imaging spectroscopy algorithms 
for partially occulted or unocculted flares. Because quantitative imaging spectroscopy of solar 
flares are limited by the low dynamical range, we only consider the imaging spectroscopy at the HXR 
peak of the two limb flares with both FPs seen and the evolution of the total 
spectrum, assuming that the impulsive phase spectral evolution of the LT source of the three flares 
with FP sources is similar to that of the LT source of the three occulted flares.

For the August 12 flare we studied the spectra above 8 keV to avoid emission line features at 
$\sim 7$ keV. SPEX can handle these emissions in a thermal model but not for the nonthermal 
ones. For the rest of the flares we limited the spectral fitting to the above $10$ keV energy 
range because at least one of the shutters was in for these flares, which results in low 
count rates and a non-diagonal detector response matrix at low energies (below 12 keV). This 
causes significant uncertainties in the spectral fitting. In many cases, both a thermal and a 
power-law model can fit the observations with similar reduced $\chi^2$.

The left panel of Figures \ref{specs.ps}-\ref{specs3.ps} show the best-fit spectra of the two 
limb flares of September 20 and August 12, and the occulted flare b of April 4 at the preheating 
phase (lower) and HXR peak (upper). For the limb flares, the high-energy HXR emission in the 
impulsive phase mostly comes from the FPs. The right panels are for an early (upper) and a late 
(lower) period in the gradual decay phase. The bottom panel of each figure gives the residual for 
the upper spectrum in unit of the standard deviation. These four periods are indicated by the 
arrows in the top panels of Figures \ref{cooling.ps}-\ref{cooling2.ps}. With {\it RHESSI}'s 
superior spectral resolution and broad energy coverage, it can be shown clearly that although the 
power-law component can still be seen early in the decay phase, the spectrum becomes purely 
thermal later on. The observations then reveal a hot plasma confined at the LT, and the 
confinement is evident at the highest energy. It is obvious that a combined study of images at 
different energy ranges will disclose the temperature and density profiles of the flare loop in 
more detail, providing critical constraints on its dynamical evolution, especially the heating and 
cooling processes. Such kind of investigations are clearly warranted.

\begin{figure}[thb]
\epsscale{1}
\centerline{\plottwo{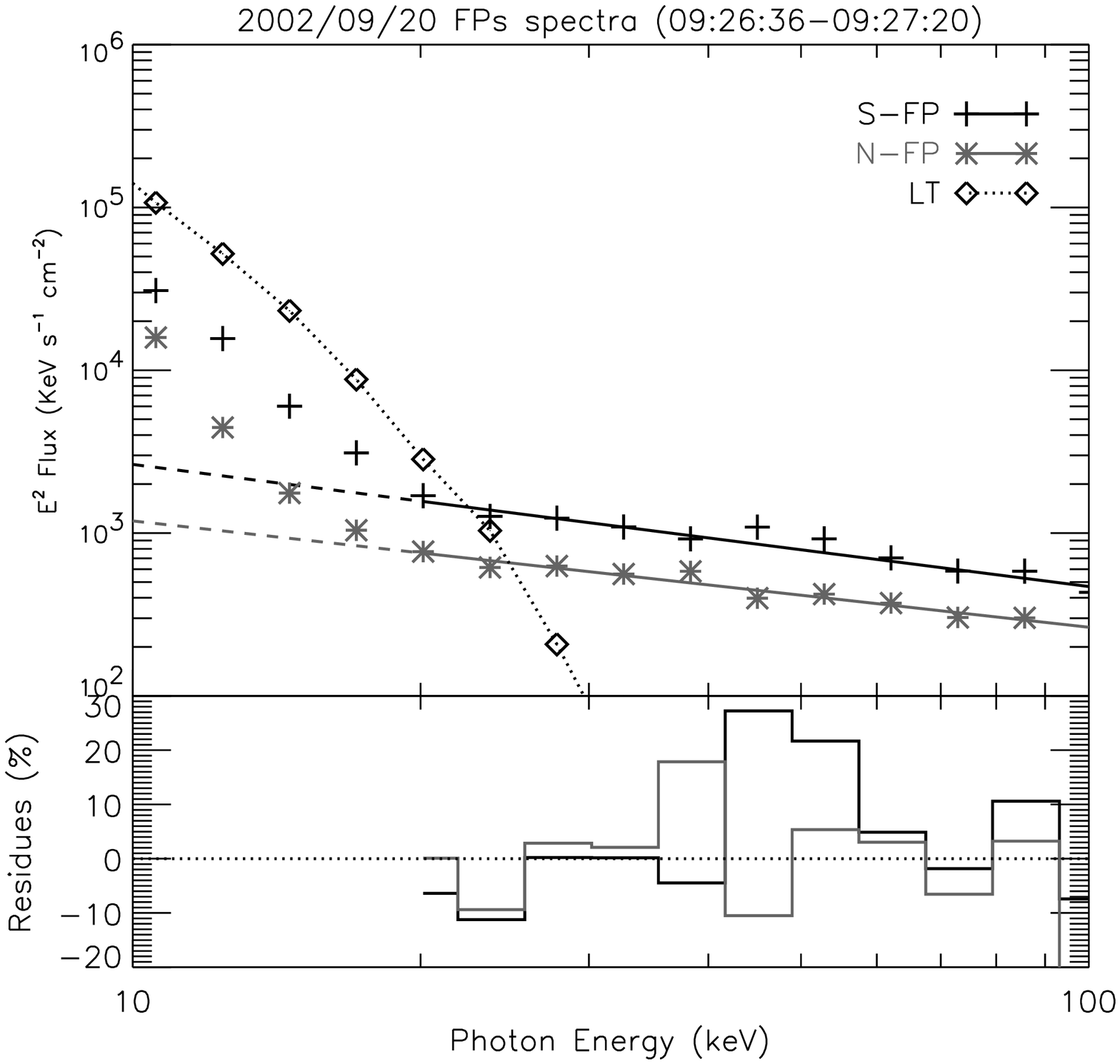}{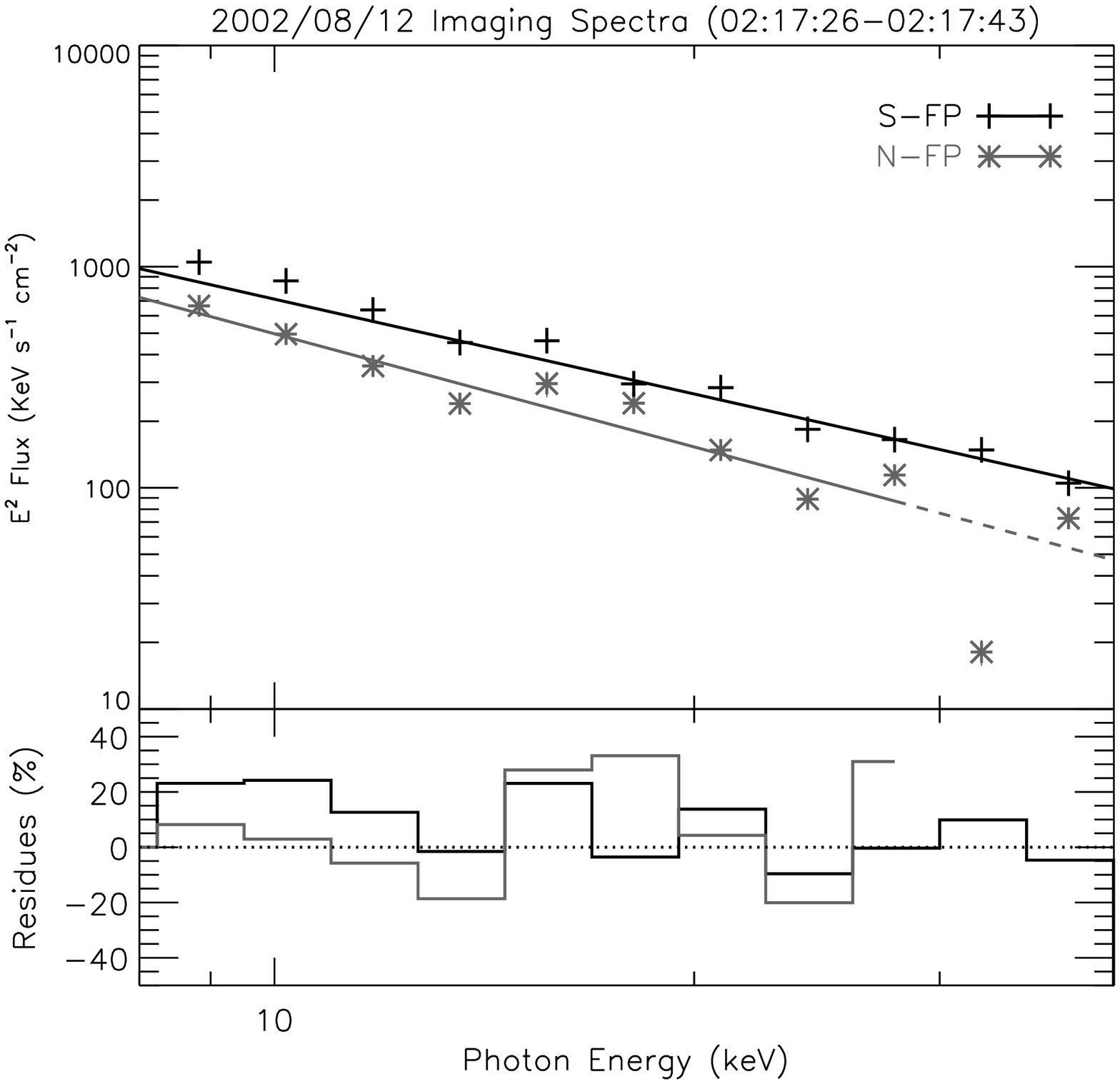}}
\caption{
Spectra of the LT and FP sources of the September 20 (left panels) and August 12 (right panels) 
flares during the HXR peaks. The solid lines give the spectral fits to the FPs: a thermal plus a 
power-law model for the former and a power-law model for the latter. The LT spectrum is fitted by 
a thermal model as indicated by the dotted line. For the August 12 flare the LT source does not 
yield a reliable spectrum in the range of figure and therefore is not shown. 
The ratio of the FP and LT fluxes at lower energies ( $< 20$ keV) of the September 
20 flare are close to the dynamical range of the PIXON algorithm. The FP fluxes are thus not 
trustworthy. The lower panels show the relative error of the spectral fittings to the FPs.
}
\label{imagspecs.ps}
\end{figure}

Figure \ref{imagspecs.ps} shows the LT and FP spectra of the September 20 (left) and August 12 
(right) flares at their HXR peak. For the former only the thermal component of the LT can be 
seen, while the thermal component of the FP spectra are likely due to contaminations by the LT 
considering the compactness of the source and the limitation of the spatial resolution of the 
PIXON algorithm. Fitting the high energy spectra of the FPs with a power-law model, we obtained a 
spectral index of 2.75 and 2.65 for the southern and the northern FP, respectively, indicating 
an injection of a power-law electron spectrum into the FPs.  The bremsstrahlung yield of this 
thick target source is then well defined (see e.g. Petrosian 1973), and gives an electron 
energy flux of $\sim 2.4\times 10^{27}(E_c/10 {\rm keV})^{-1.75}$ and $\sim 0.8\times  
10^{27}(E_c/10{\rm keV})^{-1.65}$erg s$^{-1}$ for the two FPs, where $E_c$ is the low energy 
cutoff of the power-law electron distribution. These are much higher than the bolometric 
luminosity of the thermal LT source: $\sim 10^{25}$erg s$^{-1}$. For the August 12 flare the LT 
spectrum can be determined only below 10 keV and therefore is not shown here. The electron 
fluxes at the southern and northern FPs are $\sim 2.4\times10^{27}(E_c/10{\rm keV})^{-2.4}$ and 
$\sim 2.4\times 10^{27}(E_c/10{\rm keV})^{-2.7}$erg s$^{-1}$, respectively. The impulsive phase durations 
of the September 20 and August 12 flares are $\sim 100$ and $\sim 30$ s, giving a total 
energy input of $\sim 3.2\times10^{29}$ and $\sim 1.4\times 10^{29}$ erg, respectively. 

\subsection{Spectral Evolution}
\label{evo}

The evolution of a flare is usually divided into three phases: ``the pre-heating phase'', when 
the SXR emission rises monotonically without apparent HXR emission; the impulsive phase, when 
almost all of the high-energy ($> 25$ keV) HXR emission is produced, and the emission shows 
short time scale ($< 1$ sec) variations; and the decay phase, when the HXR emission vanishes 
and the SXR emission decays gradually. For some flares the pre-heating phase may not be 
obvious or be totally absent. The impulsive and decay phases, however, are common features of 
all flares. The top panels in Figures \ref{cooling.ps}-\ref{cooling3.ps} give the light curves 
of the six flares at 6-12 keV, 25-50 keV and 50-100 keV, showing these three phases.

As expected the spectral properties of a flare during these three phases are quite different. To 
study the spectral evolution, we fit the total spectrum of these flares with a thermal plus a 
high-energy power-law or a broken power-law tail (the power-law index is usually larger at higher 
energies). The evolution of the power-law spectral index(es) $\delta$ (dot-dashed), emission 
measure $EM$ (dashed), and temperature $T$ (solid) are shown in the second panel of Figures 
\ref{cooling.ps}-\ref{cooling3.ps}. To determine the spectral parameters with a reasonable 
certainty, one needs at least 4 spectral data points for each of the thermal and nonthermal 
components. Only parameters, which are well measured, are shown in the figures.\footnote{Because 
the power-law component for the LT source is always very steep (with a photon index larger than 
5), it will dominate the thermal component at very low energies during the period when both 
components are present. We consider this scenario physically unreasonable, because the nonthermal 
particles are likely accelerated from the thermal background. The power-law component should be 
truncated above the thermal energy of the background plasma. We therefore introduce a low-energy 
turnover of $\sim 16$ keV for the power-law component (below which the photon spectrum is assumed 
to be flat) so that the thermal component dominates the low-energy emission.}

In the preheating and early impulsive phases, there is no clear evidence for a thermal component 
above 10 keV; only the power-law component is seen. The power-law component vanishes early in the 
decay phase, and the thermal component often appears after the HXR peak and gradually becomes more 
and more dominant. The fact that the nonthermal component is present from the rising phase to the 
early decay phase suggests that the particle acceleration or energy release occurs over a period 
longer than the duration of the impulsive phase. There also appears to be a correlation between the 
thermal and the power-law components. The emission measure, $EM$, of the thermal component keeps 
increasing as far as the nonthermal component exists in qualitative agreement with the thick target 
model and the Neupert effect. The temperature, $T$, however, starts to decay minutes before the 
nonthermal component vanishes, qualitatively consistent with predictions of the evaporative cooling 
model (Antiochos \& Sturrock 1978). However, it is not clear whether this evaporation is driven 
predominately by the nonthermal component or by a heat flux associated with the thermal particles. 
More detailed analyses and modeling are required to address these quantitatively.

In the three flares with detected FP sources there are no signs of FP emission 
minutes after the HXR peak, when a relatively steep nonthermal component is still present in the 
total spectrum, indicating continuous particle acceleration confined to the LT region. Evidently, 
the electrons accelerated during this phase are stopped in the coronal loop by the dense thermal 
plasma formed after the HXR peak (Veronig \& Brown 2004).

\section{THEORETICAL MODELING}
\label{ana}

The most interesting finding of this study occurs in the late decay phase, when the flare 
spectra are fitted well with a simple thermal model, while images show that the LT source 
is confined within a small region and remains stable there for several minutes. 

\subsection{Cooling Processes: Theory vs Observation}

From the images and spectral analyses we obtain the evolution of the volume $V=4\pi a^2b/3$, $EM$ 
and $T$ of the LT source. Here following the standard procedure (see McTiernan et al. 1993) we 
have assumed that the line-of-sight semi-axis of the source is equal to $a$. For a volume filling 
factor equal to unity (see, however, below), one can get the mean electron density $n_e$, the 
total energy ${\cal E}$, and the pressure $P$ of the thermal LT source with the following 
equations:
\begin{eqnarray}
n_e&=&(EM/V)^{1/2}=(3EM/4\pi a^2 b)^{1/2} \,,\\
{\cal E} &=& 3 n_e k_{\rm B} T V = 3 EM \ k_{\rm B} T / n_e\,, \\
P &=& 2 n_e k_{\rm B} T\,, 
\end{eqnarray}
where, for the sake of simplicity, we have assumed a fully ionized pure hydrogen plasma (so that 
the proton density $n_p=n_e=n$) and that the electron and proton temperatures are equal. One then 
obtains the density and energy evolution and the energy decay rate $\tau_{\cal E}^{-1} \equiv 
|\dot {\cal E}/{\cal E}|$, where $\dot{\cal E}(<0)$ is the time derivative of ${\cal E}$. The 
density evolution is indicated by 
the dotted line in the second panel of Figures \ref{cooling.ps}-\ref{cooling3.ps}, and the energy 
decay rate by the crosses in the third panels. Note that for the flare of March 28, 
the energy increases occasionally in the gradual phase. The corresponding increase rate is 
indicated by the star signs.

There are three major cooling processes: expansion, radiation, and conduction. As mentioned above 
there is no obvious expansion at the edge of the LT source, and adiabatic
expansion down the legs of the arcade loop may not be important here since the pressure
equilibrium is established on a time scale of the sound travel time $\tau_{s}\sim L/c_s \sim 20
$ s, where $L \sim 10^9$ cm is the half-length of the loop and $c_s\sim 400$ km s$^{-1}$ is the sound 
speed of the thermal plasma in the loop.  This time scale is much shorter than the energy decay time 
$\tau_{\cal E}\sim 10$ mins.

For plasmas with a temperature below 3 keV, the radiative cooling is dominated by line 
emissions (Tandberg-Hanssen \& Emslie 1988) \footnote{Clearly here we are not dealing with a pure 
hydrogen plasma.}, and the radiative cooling rate can be approximated 
as
\begin{equation}
\dot{\varepsilon}_{\rm rad} =\kappa_r n_e^2 T^{-1/2}\,\ \ \ \ {\rm with} \ \ \, \
\kappa_r = 1.42\times 10^{-19} {\rm erg\ cm}^3 
{\rm s}^{-1}{\rm K}^{1/2}\,, 
\end{equation}
where all quantities (here and in what follows) are expressed in the cgs units unless specified 
otherwise. The corresponding energy decay rate
\begin{equation} 
{\tau_{\rm rad}}^{-1} \equiv
{\dot{\varepsilon}_{\rm rad} V \over {\cal E}} = 7.7 \times 10^{-4} 
{\rm s}^{-1}
\left({EM\over 10^{48} {\rm cm}^{-3}}\right)^{1/2}
\left({T\over 10^7{\rm K}}\right)^{-3/2} 
\left({a\over 5^{\prime\prime}}\right)^{-1} 
\left({b\over 5^{\prime\prime}}\right)^{-1/2}\,
\end{equation}
is represented by the dash-dotted line in the third panel of Figures 
\ref{cooling.ps}-\ref{cooling3.ps}. (At the 
distance to the Sun, $1^{\prime\prime}\simeq7.3\times 10^7$ cm.) The radiative 
cooling rate is lower than the observed energy (or temperature) decay rate by a factor of a few, and 
the ratio of the two varies with time and is different for different flares, suggesting that the 
radiative cooling could not be the dominant cooling mechanism. This conclusion is different from that 
drawn by McTiernan et al. (1993), who used the instrument SXT that operates at lower energies and is 
most sensitive to relatively colder plasmas. The {\it RHESSI} LT source could be different from 
the SXT source because the former is usually hotter than the latter by $\sim 10^7$ K. The fact 
that the {\it RHESSI} LT source is much smaller than that of the {\it Yohkoh} further supports 
this interpretation.  

The standard thermal conductive cooling rate is given by Spitzer (1962). For the LT sources 
understudy the conduction heat flux density along the loop and the corresponding cooling rate are
\begin{eqnarray}
{\cal F}_{\rm Spit} &=& \kappa_S T^{5/2} \nabla T\,\ \ \ \ {\rm with}\ \ \ \ \ 
\kappa_S=1.0\times 10^{-6} {\rm erg}\
{\rm cm}^{-1}{\rm  s}^{-1}{\rm K}^{-7/2}\,, \\
\tau_{\rm Spit}^{-1} &\equiv&
{{\cal F}_{\rm Spit} \pi a^2\over {\cal E}} \nonumber \\
&=& 3.1 \times 10^{-2} {\rm s}^{-1}
\left({EM\over 10^{48} {\rm cm}^{-3}}\right)^{-1/2}
\left({T\over 10^7{\rm K}}\right)^{5/2} 
\left({L\over 10^{\prime\prime}}\right)^{-1}
\left({a\over 5^{\prime\prime}}\right)
\left({b\over 5^{\prime\prime}}\right)^{-1/2}
\,, 
\end{eqnarray}
where 
we have used the 
approximation $\nabla T\sim T/L$ in obtaining the last expression. The thick solid line 
in the third panel of Figures \ref{cooling.ps}-\ref{cooling3.ps} show the energy decay rates 
expected from the Spitzer conductivity.  As evident the conductive cooling rate is much higher 
than the observed energy decay rate. This result qualitatively agrees with the work by McTiernan 
et al. (1993).

However, laboratory experiments and numerical studies have shown that the actual conductivity is 
smaller than that given by Spitzer when the temperature variation length scale $L\sim10^9$ cm is 
less than 30 times longer than the mean free path of the thermal electrons 
$\lambda=1.4\times10^8(T/10^7{\rm 
K})^2(n/10^{10}{\rm cm}^{-3})^{-1}$ cm (Luciani, Mora, \& Virmont 1983; Post 1956).
This is the case for typical solar flare conditions. As a result the electrons deviate 
significantly from a Maxwellian distribution, and the conduction is suppressed. This is referred 
to as the non-local conduction 
since the energy transport depends upon the global structure of the system. In principle, one 
needs to solve the Fokker-Planck equation to address this problem. Fortunately, the cooling rate 
can be approximated fairly well with the expression (Rosner, Low, \& Holzer 1986) 
\begin{equation}
{\tau_{\rm nl}}^{-1} \simeq
0.11\left(\frac{\lambda}{L}\right)^{-0.36} {\tau_{\rm Spit}}^{-1}\,.
\label{nlcool}
\end{equation}
The thin solid lines in the third panel of Figures \ref{cooling.ps}-\ref{cooling3.ps} show the 
expected cooling rate based on equation (\ref{nlcool}), which is lower than the Spitzer rate but 
still higher than the observed energy decay rate.

Takahashi \& Watanabe (2000) observed similar discrepancy between the decay rate of {\it Yohkoh} 
flares and that expected from the Spitzer conductivity and introduced a volume filling factor 
$\eta \le 1$ to resolve it. For a loop consisting of many thin filaments the combined cross 
section area $a^{\prime2}=\eta a^2$. Since $\tau_{\rm Spit}^{-1} \propto a^2/(EM V)^{1/2}\propto 
a$, this means $\tau^{-1}_{\rm Spit}\propto \eta^{1/2}$ and can be reduced for $\eta\ll 1$ or 
densities much greater than those calculated above. In this case, however, one must also consider 
the change in the radiative cooling rate $\tau_{\rm rad}^{-1} \propto (EM/V)^{1/2}\propto 
a^{-1}\propto \eta^{-1/2}$, which increases with a smaller filling factor. Thus the total expected 
cooling rate $\tau_{\rm tot}^{-1}=\tau_{\rm rad}^{-1}\eta^{-1/2}+\tau_{\rm Spit}^{-1}\eta^{1/2}$, 
which now has a minimum of $2(\tau_{\rm Spit}\tau_{\rm rad})^{-1/2}$ at $\eta_{\rm cr} = \tau_{\rm 
Spit}^{-1}/\tau_{\rm rad}^{-1}$. With more accurate temperature and emission measure obtained with {\it 
RHESSI} we can calculate this minimum for each flares. The ratio of these rates to the observed 
cooling rate $\tau_{\cal E}^{-1}$ and the corresponding $\eta_{\rm cr}$ are shown in Table 1 for 
a typical time in the gradual phase of the six flares. As 
evident the minimum total cooling rate is in general at least one order of magnitude 
higher than the observed rate, and a low ($\lesssim 0.01$) filling factor is required, implying a 
density on the order of $10^{13}$ cm$^{-3}$ and a gas pressure comparable to or even higher than 
the magnetic field pressure. Both of these are unlikely to be the case in solar flares. We 
therefore conclude that a continuous heating (well after the nonthermal component vanishes) and/or 
a suppression of the conduction in the LT source are required to explain the observed low 
energy decay rate. In what follow we set $\eta =1$ unless specified otherwise.

\subsection{Models with the Spitzer Conductivity}
\label{uniform}

As shown in the previous section, the flare decay time is much longer than the microscopic and 
hydrodynamic times, the plasma in the loop reaches an isobaric quasi-steady state. Without the 
suppression of conduction the LT source must be continuously heated at a rate comparable to the 
conduction rate to explain the observed energy decay. In this case, however, we expect a nearly 
uniform emission flux along the corona loop, which disagrees with the observations discussed in 
\S\ \ref{img}, where the thermal source was found localized within a small region at the top of 
the flare loop.

\begin{figure}[thb]
\epsscale{0.9}
\centerline{
\plottwo{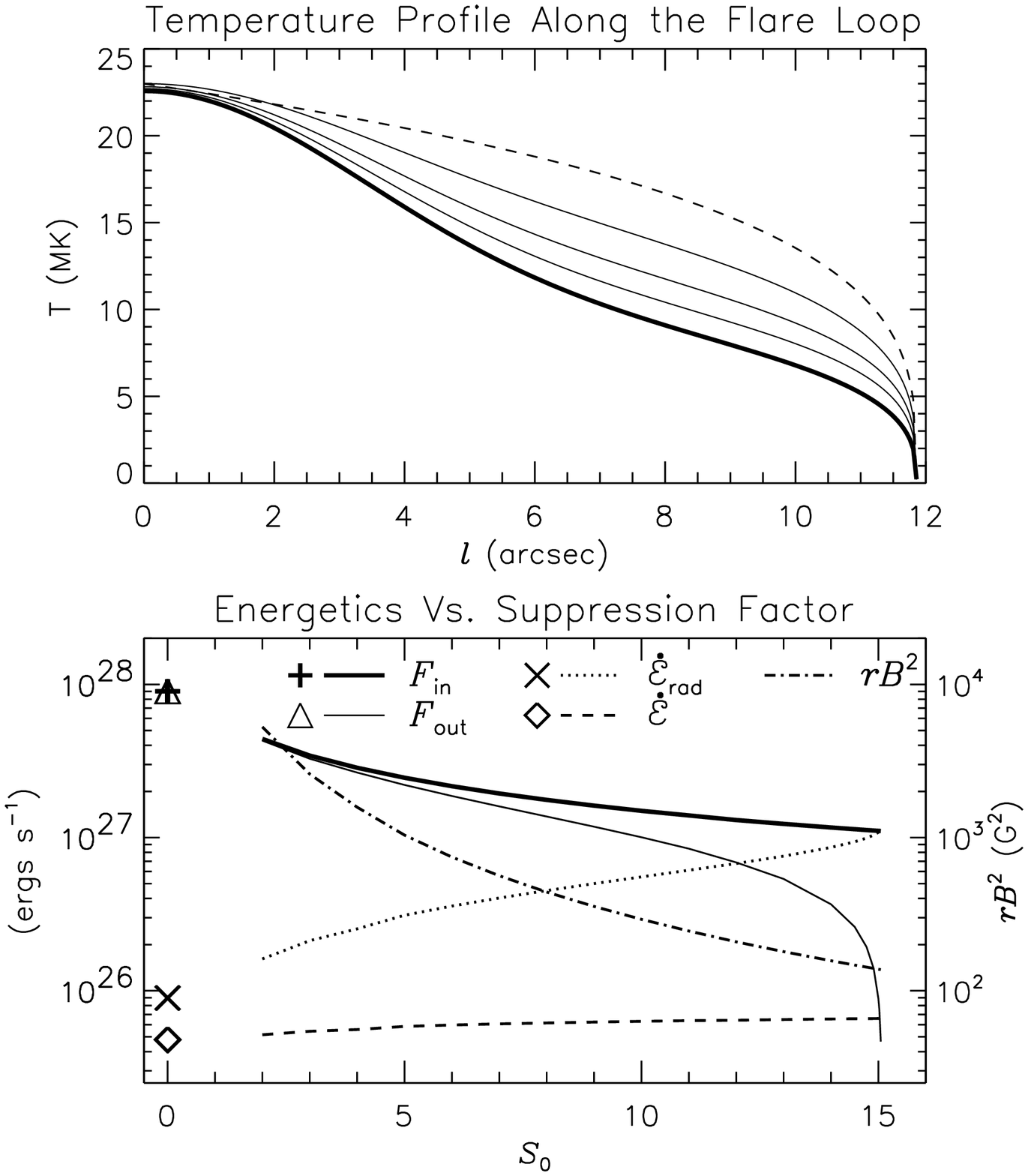}{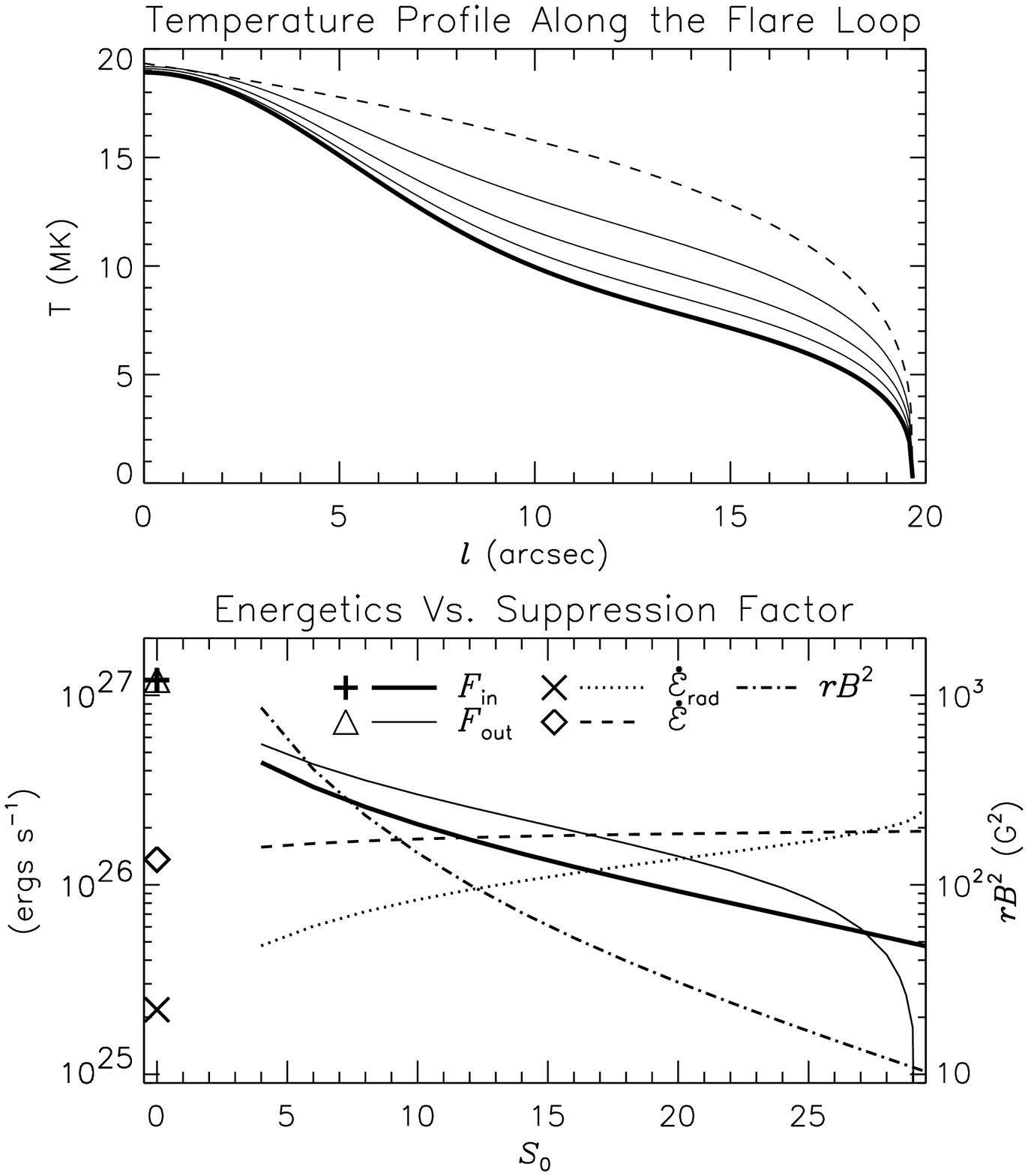}
}
\caption{
{\it Left}: The temperature profiles (top panel) for several isobaric static loop models. The top 
dashed line is for a model without suppression of conduction and with a heat flux injected at the 
LT $F_{\rm in}$. For the rest profiles turbulence is assumed to distribute at the LT,  and the 
ratio of scattering rates of the electrons by the waves and background particles $\tau_{\rm 
sc}^{-1}/\tau_{\rm Coul}^{-1}$ is proportional to $T^{5/2}$ and has a Gaussian form with a width 
$w=4\farcs5$ and a peak value of $S_0$ along the loop. The top panel shows how the temperature 
profile changes with $S_0$, which is, respectively, 15, 9, 5, 2 for the solid lines from bottom to 
top. The model shown by the thick solid line is for the maximum possible conduction suppression 
(see text) and is then used for imaging simulation 
in Figure \ref{model1.ps} and \ref{model2.ps}. The lower panel shows how the energetics changes 
with $S_0$. Here the total heating rate by the waves $F_{\rm in}= \int3 n_e k_{\rm B} T <\tau_{\rm 
ac}^{-1}>{\rm d}V$, the radiative cooling rate $\dot{\cal E}_{\rm rad} = 
\int\dot{\varepsilon}_{\rm rad}{\rm d}V$, the observed energy decay rate $\dot{\cal E}$ and the 
conduction heat flux at FPs $F_{\rm out}$ are designated by the thick solid, dotted, dashed and 
thin solid lines, (and by the plus, cross, diamond, and triangle signs for the model without a conduction 
suppression,) respectively. {\it 
Right}: Same as the {\it Left} but for the flare on August 12 
2002. Here $w=6\farcs5$, and in the upper panel $S_{0}=4, 10, 18, 28$ for the solid lines from top 
to bottom. 
}
\label{profile.ps}
\end{figure}

To quantify this study, we consider a loop model with a heat flux injection confined to the very 
top of the loop (not along the loop). The LT temperature $T_{LT}$ is determined from the spectral 
fitting, and those at the FPs are chosen to match the chromospheric value $T_{FP}\sim 0.2$ MK. (We 
do not attempt to include the even colder photosphere region because other effects, such as 
convective turbulence, will make our model inapplicable. See e.g. Spicer 1979.) The 
magnitude of the heat flux clearly depends on the conduction. The gravitational effects can be 
negligible because of the high gas pressure. With the radiative and 
conductive cooling processes, and the observed energy decay rate $\dot{\cal E}$ taken into 
account, we have the energy conservation equation within the loop
\begin{equation}
\frac{\d (a^2{\cal F}_{\rm cond})}{a^2\d l} = -\dot{\varepsilon}_{\rm rad}(l)-\dot{\cal E}/V\,,
\label{eng1}
\end{equation}
where $l$ is the distance along the loop from the middle of the LT source, and ${\cal F}_{\rm 
cond}$ gives the heat flux 
density due to conduction. Since the loop is in pressure equilibrium, $\dot{\cal E}$ is 
independent of $l$. For loops with constant cross section (i.e., $a$ independent of $l$), we get
\begin{equation}
\frac{\d}{\d l} \left(S^{-1}\kappa_ST^{5/2} \frac{\d T}{\d l} \right)
= - \kappa_r n_e^2T^{-1/2} -\dot{\cal E}/V\,,
\label{cond}
\end{equation}
where $S$ is the conduction suppression factor. Solving this equation with $S=1$, we obtain the
temperature profile along the loop shown by the dashed lines in the upper panels 
of Figure \ref{profile.ps} for the flares on September 20 (Left panel: $n_e =8\times 
10^{10}$cm$^{-3}$, $T_{LT}=2.3\times10^7$ K, $L=11.5^{\prime\prime}$) and August 12 (Right panel: 
$n_e=4\times 10^{10}$cm$^{-3}$, $T_{LT}=1.9\times10^7$ K, $L=19.5^{\prime\prime}$), where the 
physical conditions are for the periods right after the power-law component vanishes.
Since a nearly isothermal heating region will make the LT source even larger than the observed 
source size, we ignore the size of the heating region (see discussion below). As evident, in this 
case the expected radiation along the loop will be nearly uniform and not as highly confined to 
the LT region as observed in the {\it RHESSI} band.

We are mostly interested in the brightness profile along the loop, and the above analyses give 
only this variation. The source structure in perpendicular to the loop (and the field line) is not 
constrained by these analyses. In order to compare the model with observations we convolve the 
model predicted emission from the one dimension loop with a two dimensional Gaussian profile with 
its width determined by the observed semi-minor axis $a$. We simulate such a loop structure as it 
would appear in the $6-12$ keV band when observed with {\it RHESSI} and analyzed with the PIXON 
algorithm. The blue curves in Figure \ref{model1.ps} show the 15\% and 75\% contours of the 
simulated images. It is evident that the model predicted emission extends further towards the FPs 
than the observed emission (red contours). Figure \ref{model2.ps} compares the model predicted 
brightness profiles (dashed lines) with observations. The agreement in the perpendicular direction 
is a result of our convolution (lower panels). But along the loop (upper panels), which is what 
matters here, the LT source is extended with an FWHM $\sim 30$\% and 50\% longer than that 
observed for the September 20 (left) and August 12 (right) flares, respectively. These are longer 
than the observed source lengths by about one FWHM and one and a half FWHMs of the PSF (shown as 
dot-dashed profiles), respectively.

The lower panels of Figure \ref{profile.ps} show the energy fluxes $F_{\rm in}$ and 
$F_{\rm out}$ injected at the LT (plus) and the FPs (triangle), respectively, the total radiative 
loss $|\dot{\cal E}_{\rm rad}|=\int\dot\varepsilon_{\rm rad} {\rm d}V$ (cross), and $|\dot{\cal E}|$ 
(diamond). The energy balance is achieved here by introducing a high heating rate at the LT, most 
of which is conducted to the FPs. $F_{\rm  in}\simeq F_{\rm out}\sim 10^{28}$, and $10^{27}$ erg 
s$^{-1}$ for the September 20 and August 12 flares, respectively. These are almost two order of 
magnitude higher than the respective radiative loss rates. There are several difficulties with 
this scenario. The most important one is that this high rate (over several minutes) implies a 
total energy input $>10^{30}$ erg for the September 20 flare (with a duration $>100$ s), which is 
much larger than the energy input of $\sim 3\times 10^{29}$ erg estimated in \S\ \ref{spec} for the 
impulsive phase. If this energy is converted into radiation at the FPs, there will 
be significant radiative signatures (Spicer 1979), which is not observed. Most of this energy 
therefore has to be deposited into the cold chromospheric plasma and would cause evaporation, 
which appears to contradict the observed slow decrease of the $EM$ during this phase. One possible 
(but unlikely) scenario is that most of the plasma evaporated from the FPs has a very low 
temperature and does not contributed to the observed X-ray emission. Detailed 
analyses of GOES observations of these flares may shed light on this problem (Allred et al. 2005).

With a volume filling factor $\eta\ll 1$,  the conductive heat flux can be reduced. However, 
this will not alter the temperature profile significantly, so that the discrepancy with the 
confinement of the LT remains unresolved. An upward flow of evaporated plasma from the FPs can 
reduce the conductivity, but it also will not modify the temperature profile of the loop 
significantly 
(Antiochos \& Sturrock 1978). 
The recently proposed multiple-loop models for solar flares may explain the discrepancies in 
the cooling rates, since the observed energy decay is controlled by the assumed loop formation 
rate (Reeves \& Warren 2002; Warren \& Docshek 2005). However, this model also can not explain 
the confinement of the LT source. Therefore all these effects may account for the 
discrepancy in the cooling rates and reduce the apparent size of the LT but fail to explain the 
observed strong confinement due to the nearly uniform temperature profile resulting from the 
sensitive dependence of the thermal conduction on the temperature.
We conclude that in the late decay phase these observations suggest that the conductivity 
has to be suppressed significantly.

\subsection{More Realistic Models with Suppressed Conductivity and Heating}

The above discrepancies between the observations and the simple model, primarily due to the
confinement of the LT source, can be overcome with a suppression of conductivity. As we will
see below even in this case the LT source must be heated continuously (but at a slower rate).
We propose that  plasma wave and turbulence produced at the LT can be responsible for both
these effects. In this section we investigate the characteristics of such a scenario.

We first point out that a uniform suppression of conductivity along the whole loop cannot
resolve these discrepancies. We have solved equation  (\ref{cond}) for various values of the 
suppression factor $S>1$ and found that the temperature profile changes little with $S$. This is 
because the conductive heating flux and radiative cooling rate are very sensitive to the 
temperature, the temperature has to be nearly uniform to satisfy the boundary conditions.
When the thermal conduction dominates the loop must be nearly uniform to carry an almost constant 
heat flux for given temperatures at the LT and FPs. When the radiative cooling becomes important, 
($\dot\epsilon_{\rm rad}\propto n^2T^{-1/2} \propto T^{-5/2}$ for the isobaric model and for the 
range of temperature of interest here), the cooling rate is also very sensitive to temperature, 
and the plasma is in an unstable state. The loop temperature again has to be nearly uniform to 
satisfy the boundary conditions at the FPs\footnote{In principle, one can prescribe a heating 
function along the loop so that the radiative cooling process can be balanced by the divergence 
of the conductive heat flux, the observed energy decay and the local heating (Rosner, Tucker, \& Vaiana 
1978). A steep temperature profile in consistence with the observed LT confinement might be realized if 
the heating function were fine-tuned. We consider this scenario unphysical and will ignore it in the 
following discussion.}.

Therefore we consider a model, where the suppression and heating (and the turbulence responsible for them) 
are confined to the LT region. Roughly speaking the loop consists of two parts:
a hot LT region where the observed source locates and cold legs of comparable size.

We have compared with observations many simulated images with different temperature profiles as an 
input and found that the temperature of the plasma near the $15\%$ contour 
of the observed images has to be at least two times lower than the measured source temperature. 
We first note that as a consequence of the isobaric assumption this means that independent of the 
details of the model heating may be required because the radiative cooling rate will be at 
least $2^{5/2}\simeq 6$ times higher in the legs than in the hot LT source. For the September 20 
flare, this will exceed the observed energy decay rate.  The conductive cooling rate ($\propto 
T^{5/2}$) will be $>6$ times lower (here we assume that the mean temperature gradient for the loop 
does not change), but can still be the dominant cooling process, implying a suppression of conduction in 
the LT region by at least a factor of 6. Note that the ratio of conductive to radiative cooling 
rates is proportional to $T^5$ so that if the temperature of the relatively cold legs  
$T_{Leg}<T_{LT}(\tau_{\rm Spit}/\tau_{\rm rad})^{1/5}$ the radiative cooling in the leg will 
exceed the available heating carried by conduction from the LT region. The requirement of 
$T_{Leg}<2T_{LT}$ then implies that $\tau_{\rm rad}/\tau_{\rm Spit}$ must be larger than $\sim 32$ 
for the loop model to be applicable. For the six flares understudy this requirement is satisfied. 
(It is marginal true for the partially occulted flare ``a'' on April 4). The weak dependence of 
this limit on the ratio of the observed conductive and radiative cooling rates also implies that 
the temperature in the legs can not be more than $\sim3$ times lower than 
the measured temperature of the LT source (Table 1).

In summary we require a suppression of conductivity and heating in the LT source but the heating  
rate must be lower than the value obtained in the previous section, where the conduction is not
suppressed.

A continuous but slow production of plasma waves or turbulence via a relatively slow magnetic   
reconnection above the LT in the decay phase may explain the localized suppression of conduction
and heating in the LT region.  In such a scenario the conductivity is suppressed due to the
scattering of high-energy particles in the hot plasma by the waves, which reduces the particle
mean-free-path or scattering time and suppresses conduction. The dissipation of the wave energy,
on the other hand, can result in heating of the plasma. Here we consider a simple case, where the
scattering rate of particles by the waves $\tau_{\rm sc}^{-1}$ is independent of the particle
velocity $v$ (and therefore the temperature $T$) but is proportional to the turbulence energy 
density. The corresponding conductive heating flux is given by (Spicer 1979):
\begin{equation}
{\cal F}_{\rm cond} =
{1\over 1+\tau_{\rm Coul}/\tau_{\rm sc}}
{\cal F}_{\rm Spit}= \left({\kappa_{S} T^{5/2}\over 1+S(l)T^{5/2}} \right) \nabla T\
\,,
\end{equation}
where $\tau_{\rm Coul}^{-1} \simeq 15 (T/10^7{\rm K})^{-3/2}(n/10^{10}{\rm cm}^{-3})$ s$^{-1}$ is
the mean Coulomb collision rate of the thermal electrons carrying the heat flux, and the ratio of
the mean wave scattering to coulomb collision rates is given by $\tau_{\rm sc}^{-1}/\tau_{\rm
Coul}^{-1}=S(l) T^{5/2}$ (assuming isobaric condition).
The suppression factor $S$ will be proportional to the wave energy
density and will be expected to decrease toward the FPs. In what follows we describe the spatial 
variation of the suppression by a Gaussian function of width $w$ centered at the centroid of the 
LT source, i.e. $S=S_{0} \exp\left(-l^2/w^2\right)T_{LT}^{-5/2}$.
If the waves and particles are coupled via a resonance process, the corresponding particle
acceleration rate $\tau_{\rm ac}^{-1} \simeq \xi (v_{\rm A}/v)^2 \tau_{\rm sc}^{-1}$, where
$v_{\rm A}=B/\sqrt{4\pi nm_p}$ is the Alfv\'en velocity ($B$ is the magnetic field, $m_p$ proton 
mass), and the coefficient $\xi$ depends on the wave spectrum and
details of the wave-particle coupling (Schlickeiser 1989). The energy conservation equation
(\ref{eng1}) is thus modified to:
\begin{equation}
\frac{\d (a^2{\cal F}_{\rm cond})}{a^2\d l} = -\dot{\varepsilon}_{\rm rad}(l)-{\dot{\cal E}\over
V} + {{\cal E}<\tau_{\rm ac}^{-1}>\over V}\,,
\label{eng2}
\end{equation}
where $<\tau_{\rm ac}^{-1}>$, the average of the acceleration rate $\tau_{\rm ac}^{-1}$ over the
particle distribution, denotes the heating rate. For a loop with a constant cross section, this
equation becomes
\begin{equation}
\frac{\d}{\d l} \left({\kappa_S T^{5/2}\over 1 +S(l) T^{5/2}}
\frac{\d T}{\d l} \right) = - \kappa_r n_e^2T^{-1/2} + 3n_e k_{\rm B} T\left[\tau_{\cal
E}^{-1} + 3\xi (v_{\rm A}/v_{\rm th})^2 S(l) T^{5/2} \tau_{\rm Coul}^{-1}\right]\,,
\end{equation}
where $v_{\rm th}=(3k_{\rm B}T/m_e)^{1/2}$ is the thermal velocity of the electrons and $m_e$ is
the electron mass. Because heating is distributed throughout the LT region, one can set the heat
flux at the top of the loop, i.e. at $l=0$, equal to zero, which gives a boundary condition. We
keep the other boundary conditions of the LT and FP temperatures the same as those in the previous
section. With such a model the temperature profile can be obtained once one specifies $S_0$ and
$w$. However, to satisfy the boundary conditions at the LT and FPs, the parameter $\xi$ has to
attain certain specific values, implying that heating is required. This is essentially an eigenvalue 
problem with $\xi$ being the
eigenvalue and $T(l)$ the eigenfunction.

For the physical conditions used before for the flares on September 20 and August 12, in Figure  
\ref{profile.ps} the solid lines in the upper panels show several of these profiles with the
corresponding $S_0$ and $w$ indicated in the caption. The lower panels show the dependence of the
loop energetics on $S_0$, where $F_{\rm in}$ (thick solid), $F_{\rm out}$ (thin solid), $\dot{\cal
E}_{\rm rad}$ (dotted), and $\dot{\cal E}$ (dashed) give the spatially integrated heating by the
waves, the heating flux injected into the FPs, the spatially integrated radiative cooling, and the
observed energy decay, respectively. The dot-dashed line shows the dependence of $\xi B^2$ on
$S_0$ with the scale indicated on the right frame. As expected $F_{\rm in}$ and $F_{\rm
out}$ decrease with $S_0$. These are consequences of the conduction suppression, which reduces  
not only the level of heating needed to keep the energy balance but also the conductive heat flux
injected into the FPs. The parameter $\xi B^2$, therefore, decreases with $S_0$. The radiative
cooling rate, on the other hand, increases with $S_0$ because the overall temperature of the loop
decreases (see the top panel). As expected the observed energy decay rate only weakly
depends on $S_0$.

It is clear that the energy balance is achieved by the interplays of the conduction, radiation,  
heating, and the observed energy decay. For the September 20 flare the heating process always 
dominates and is balanced by conduction and radiative cooling with the latter being more important
with the increase of $S_0$. The observed energy decay is slow and therefore not important. These
are quite different from those of the August 12 flare, where the observed energy decay dominates at large 
values of $S_0$, and for small values of $S_0$ the conductive cooling dominates. Interestingly,
for both flares $S_0$ has a maximum, where $F_{\rm out}=0$, indicating no chromospheric
evaporation in the phase, which is in line with the slow changes or even decay of the observed
$EM$ (Fig. \ref{cooling.ps}). Above this maximum, no physical solution exist because conduction is
not efficient enough to balance the rapid radiative cooling near the FPs. The models with the
maximum conduction suppression at the LT are indicated by the thick solid lines in the upper 
panels. The simulated images for these models are indicated by the green contours and dotted lines 
in Figures \ref{model1.ps} and \ref{model2.ps}, which agree with the observed LT confinements.    

\begin{figure}[thb]
\epsscale{1}
\centerline{
\plottwo{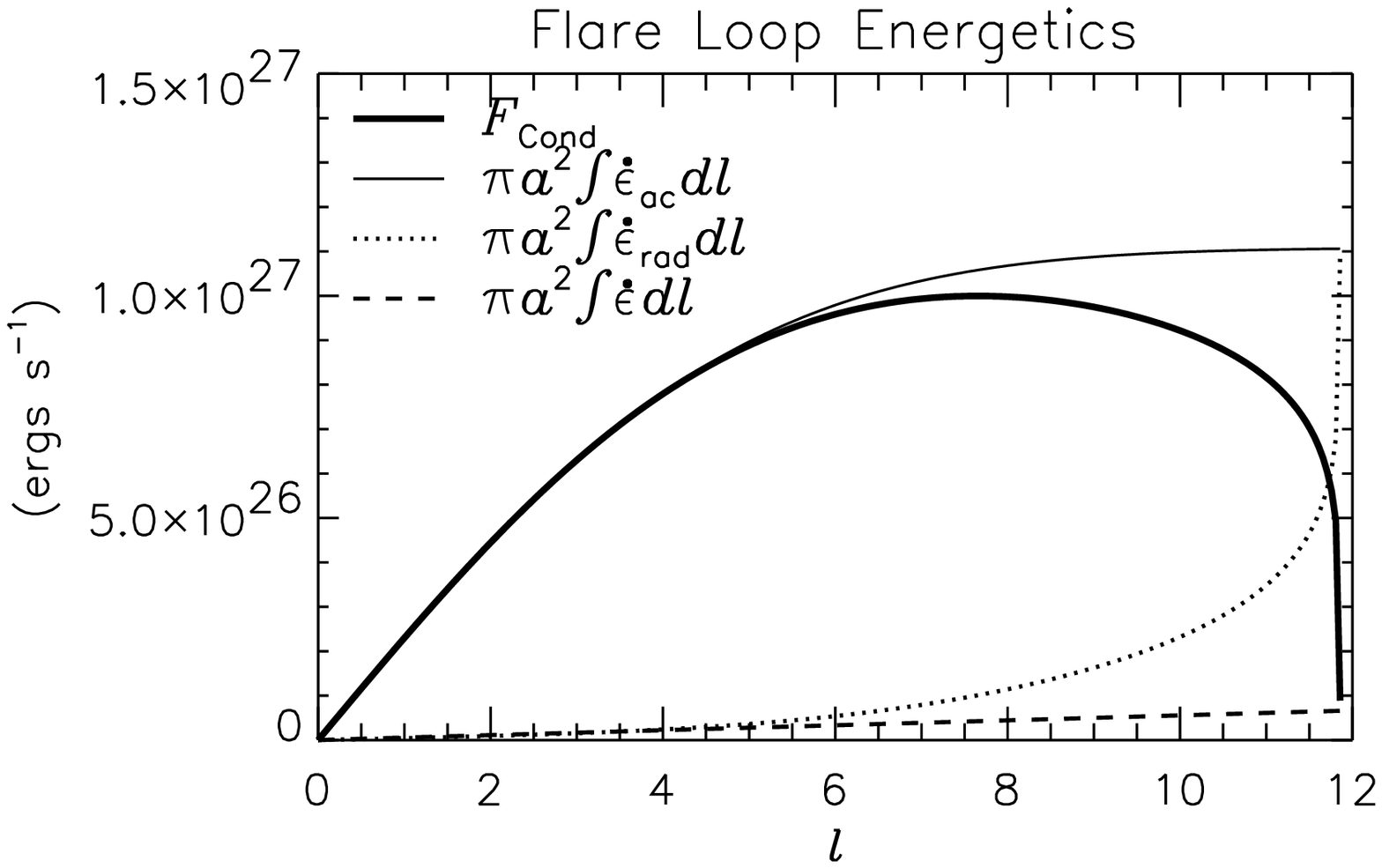}{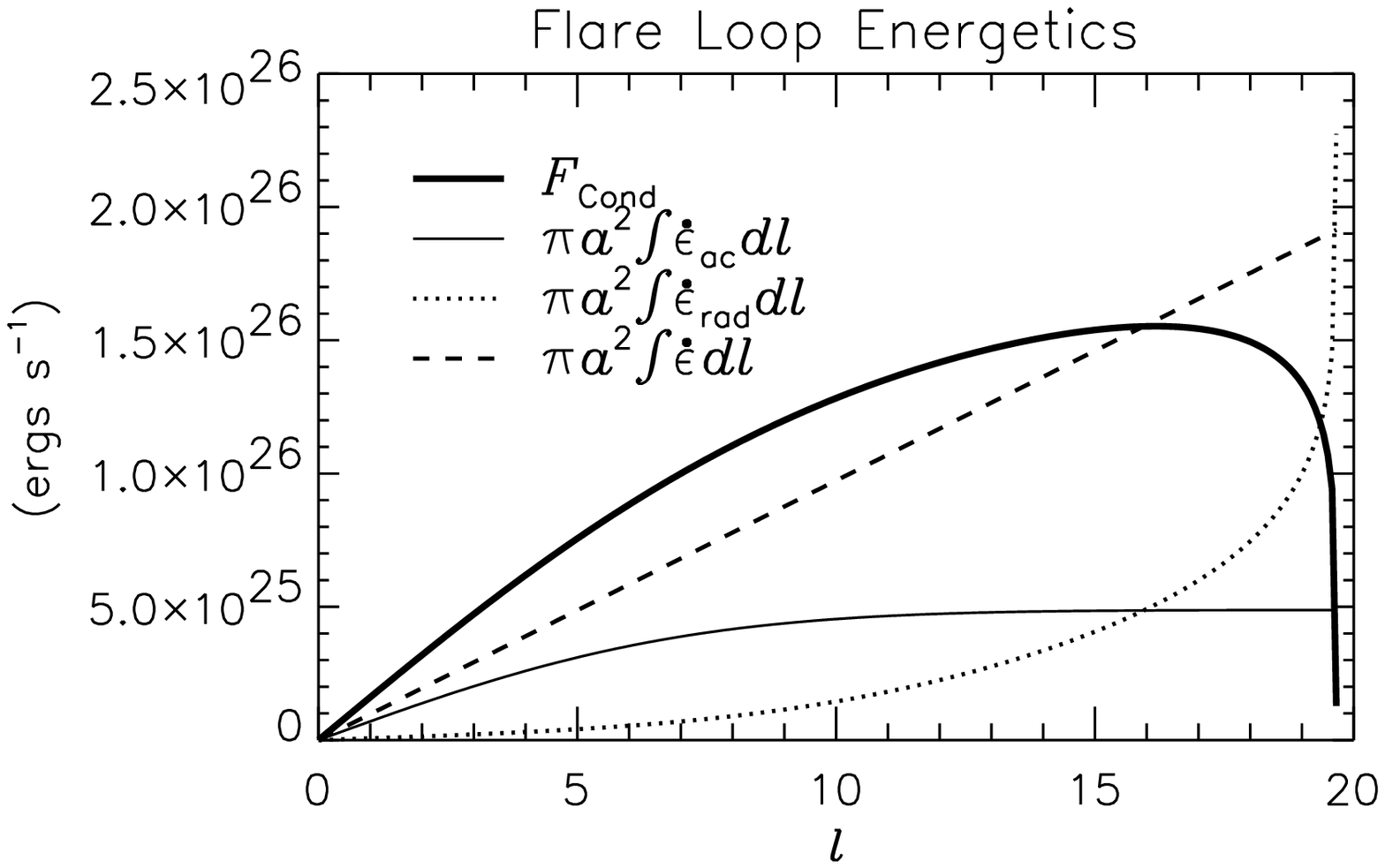}
}
\caption{
{\it Left}: The energetics along the loop of the model with a maximum conduction suppression for 
the September 20 flare. The conductive heating flux is indicated by the thick solid line. The thin 
solid, dotted, and dashed lines give the heating, radiative cooling, and energy decay rates 
integrated from the LT, respectively. Note that more than half of the radiative energy is carried 
away near the FPs. 
{\it Right}: Same as the {\it Left} but for the flare on August 12 2002. Heating is less important 
here, and the observed energy decay can be attributed to cooling near the FPs. 
}
\label{profileeng.ps}
\end{figure}

These models therefore account for the observed flare decay. Because there are no heat injections 
at the FPs, the radiative cooling process plays a dominant role then. To further illustrate the 
details of the 
model, Figure \ref{profileeng.ps} shows the flare energetics along the loop. It is important to 
note that more than half of the radiation is produced near the FPs. Strong optical and UV line 
emissions are thus expected from the FPs. The model predicted bolometric luminosities (at the 
periods of interest) of the September 20 and August 12 flares are $\sim10^{27}$erg s$^{-1}$ and 
$\sim 2\times 10^{26}$erg s$^{-1}$, respectively. These predictions may be tested with 
observations over a broad energy band.

For the September 20 flare the total energy injection in the decay phase is $\sim 6\times 
10^{29}$erg (assuming a duration of $\sim 10$ mins), which is comparable to that in the impulsive phase 
(see \S\ \ref{spec}). The total energy injection in the decay phase of the August 12 flare ($\sim 10^{28}$ 
erg s$^{-1}$ for a period of $\sim 3$ mins), on the 
other hand, is more than one order of magnitude lower than that injected during the 
impulsive phase, and the flare decay is dominated by the radiative cooling process although a 
suppression of conduction is still required. The Neupert effect therefore gives a reasonable 
description of this flare. These results, specially those for the September 20 flare, suggest 
that flares with a longer decay time require more energy injection in the decay phase and are 
not in line with the Neupert relation. 

Because it is believed that the energy of solar flares comes from the magnetic field, the low 
values of $\xi B^2$ inferred from the modeling ($\sim130$ G$^2$ and $\sim 10$ G$^2$ for the 
September 20 and August 12 flares, respectively) suggest a $\xi$ much less than $1$ for the 
wave-particle interaction. This means much more efficient particle scattering than plasma heating 
by the waves. Once the wave modes responsible for these processes are identified, the observations 
can be used to constrain the wave energy density and spectrum (Schlickeiser \& Miller 1998).

\section{Summary and Conclusions}
\label{con}

\begin{figure}[thb]
\epsscale{.5}
\centerline{
\plotone{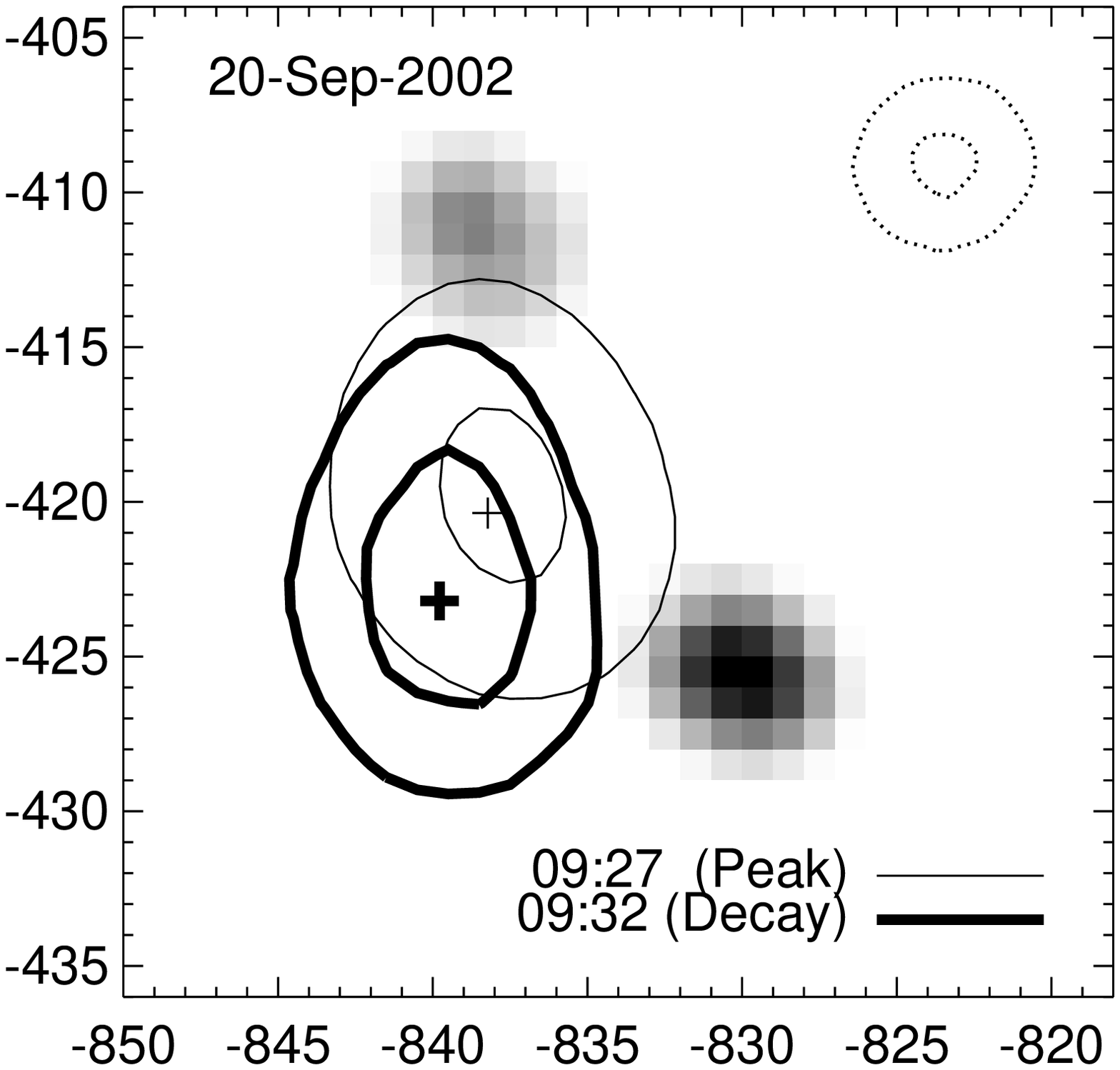}
\plotone{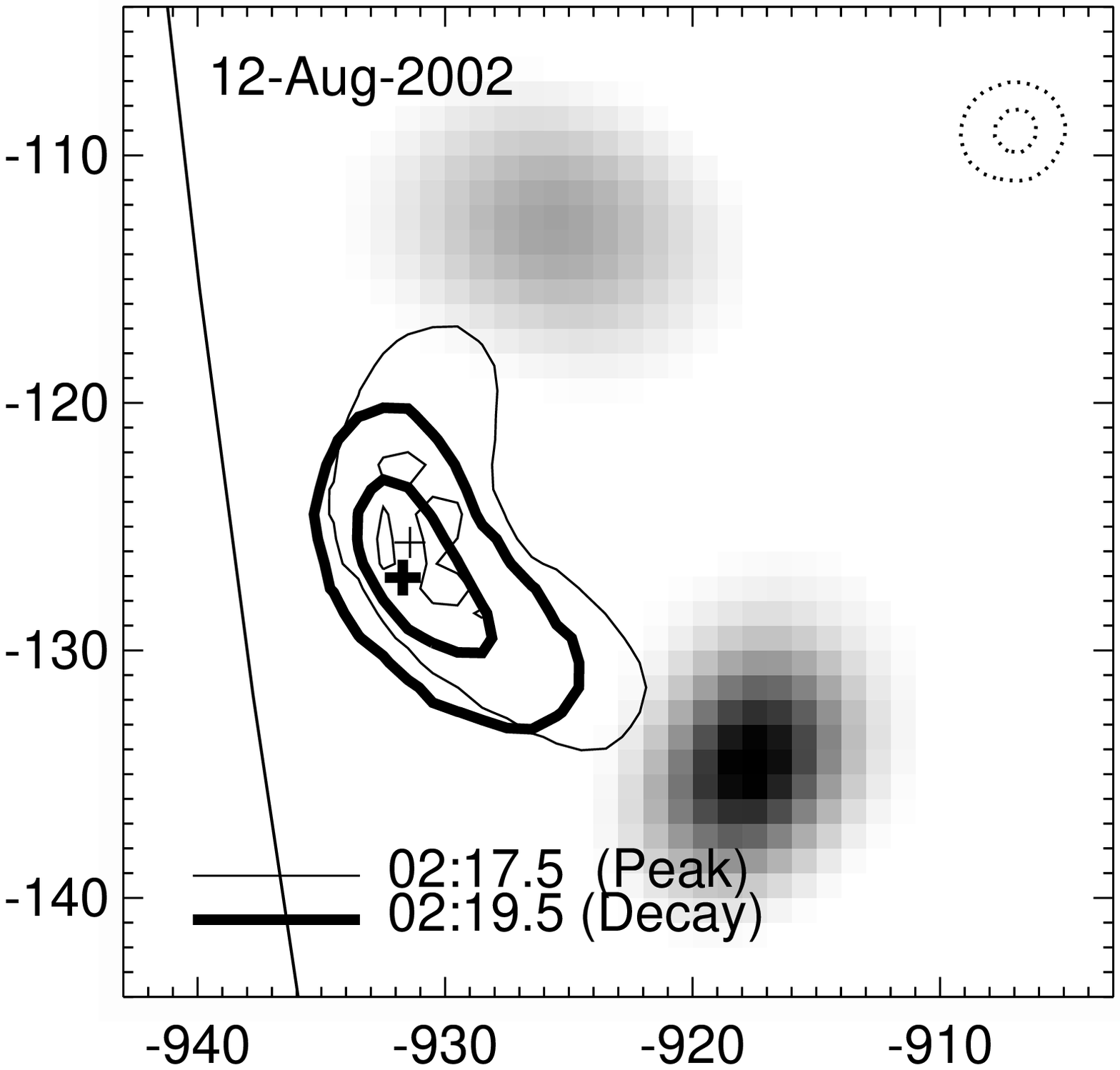}
}
\caption
{
{\it RHESSI} PIXON images of the LT source (6-12 keV) and FP sources (at 51-57 keV and 34-39 keV, 
respectively) of the two limb flares on September 20 and August 12. The thin contours (15\% and 
75\% of the peak brightness of the image, which are adopted hereafter unless specified otherwise) 
and the gray scale are for the HXR peak (see light curves in figure \ref{cooling.ps}). The thick 
contours are for a period in the decay phase. The integration time for each image is $\sim 20$ 
seconds with the corresponding start times indicated in the figure. The dotted contours at the 
upper-right corner give the PSF at 6-12 keV.
}
\label{images1.ps}
\end{figure}

\begin{figure}[thb]
\epsscale{.5}
\centerline{
\plotone{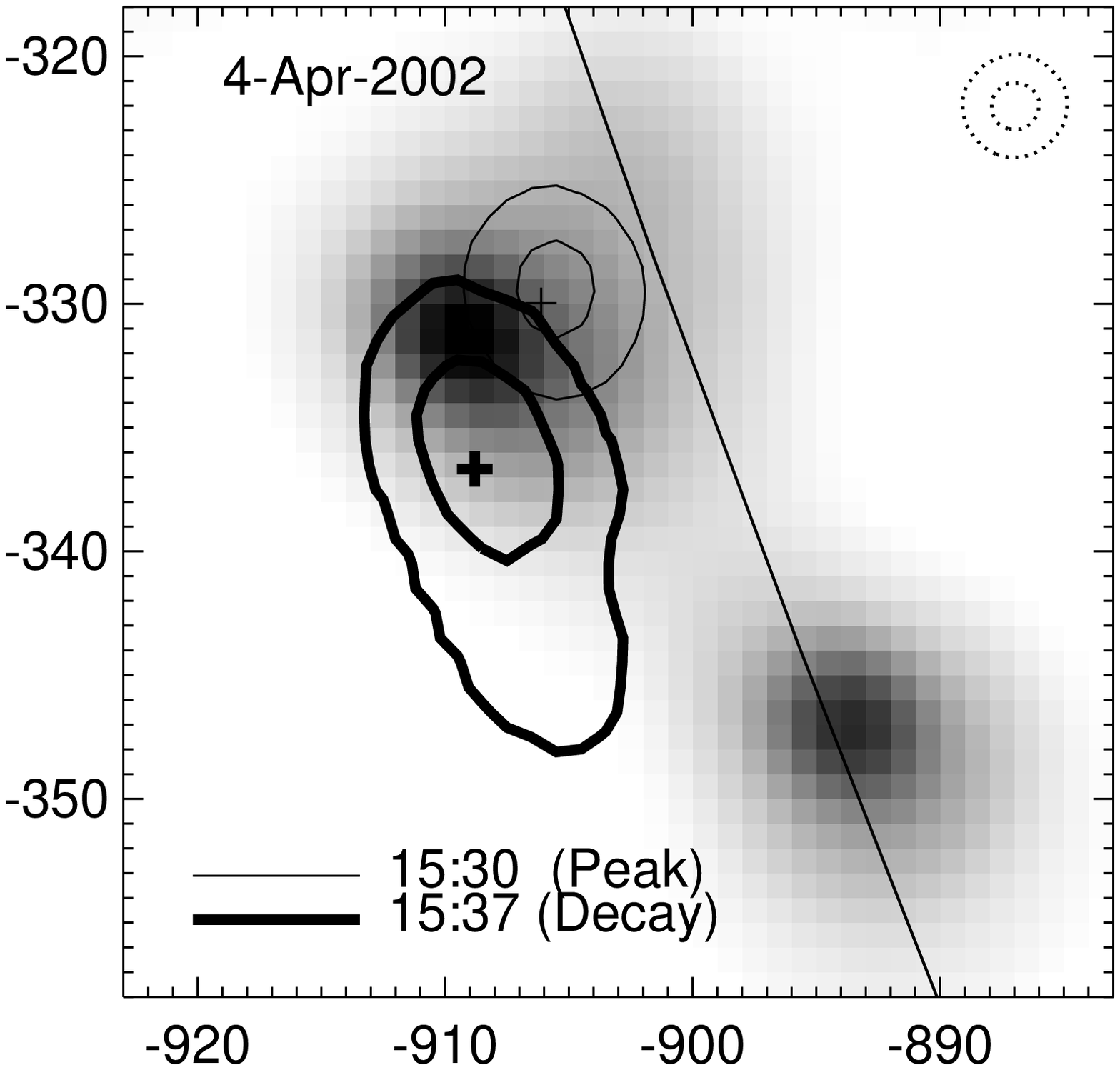}
\plotone{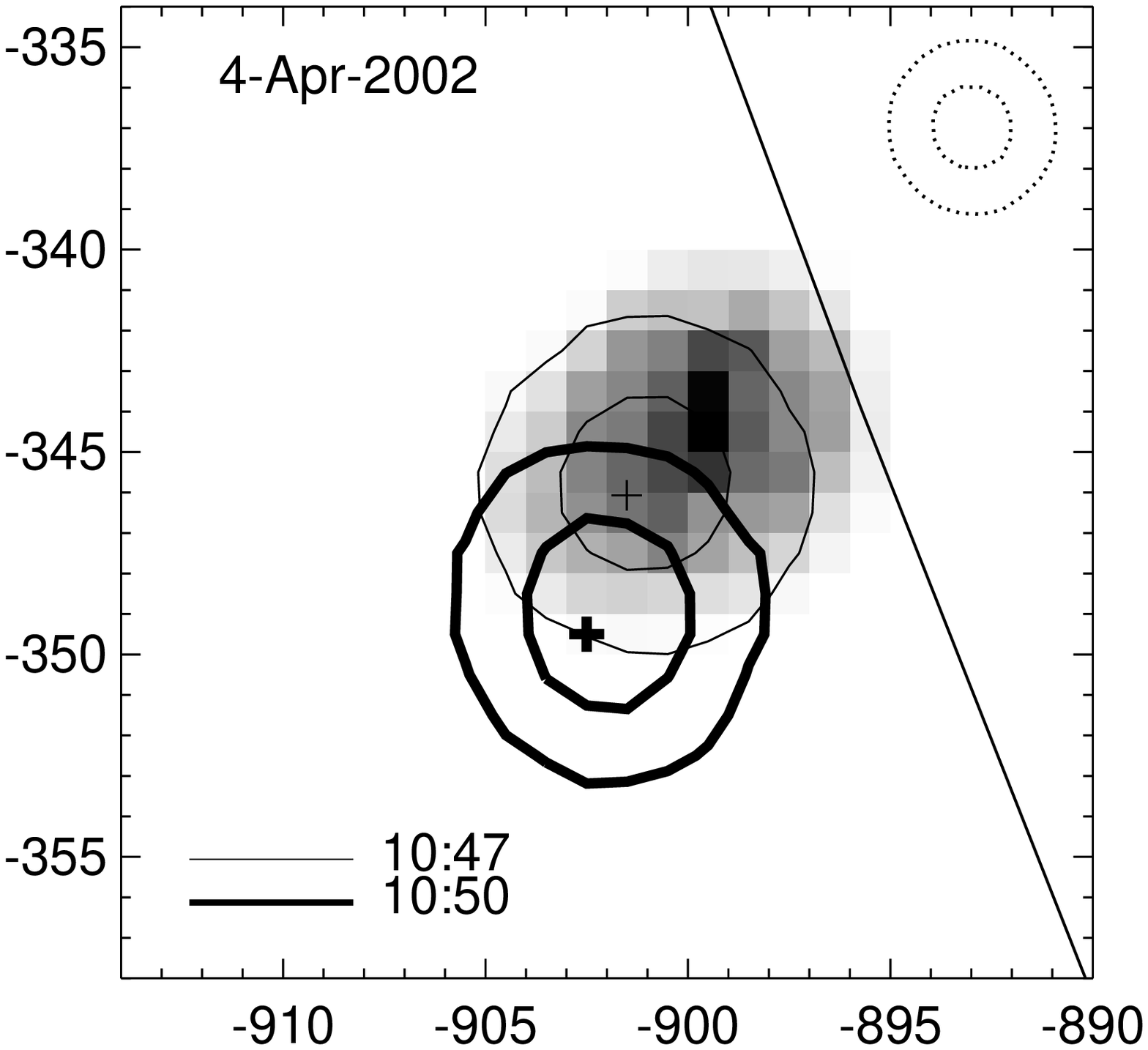}
}
\caption{
Same as Figure \ref{images1.ps} but for the partially occulted flare a and the corona event b 
on April 4. The grey scale is at 60-100 keV for the former, and for the latter the images are at 
6-12 keV with the grey scale for the HXR peak and the thin and thick contours for two periods in 
the decay phase as indicated.
}
\label{images2.ps}
\end{figure}

\begin{figure}[thb]
\epsscale{.5}
\centerline{
\plotone{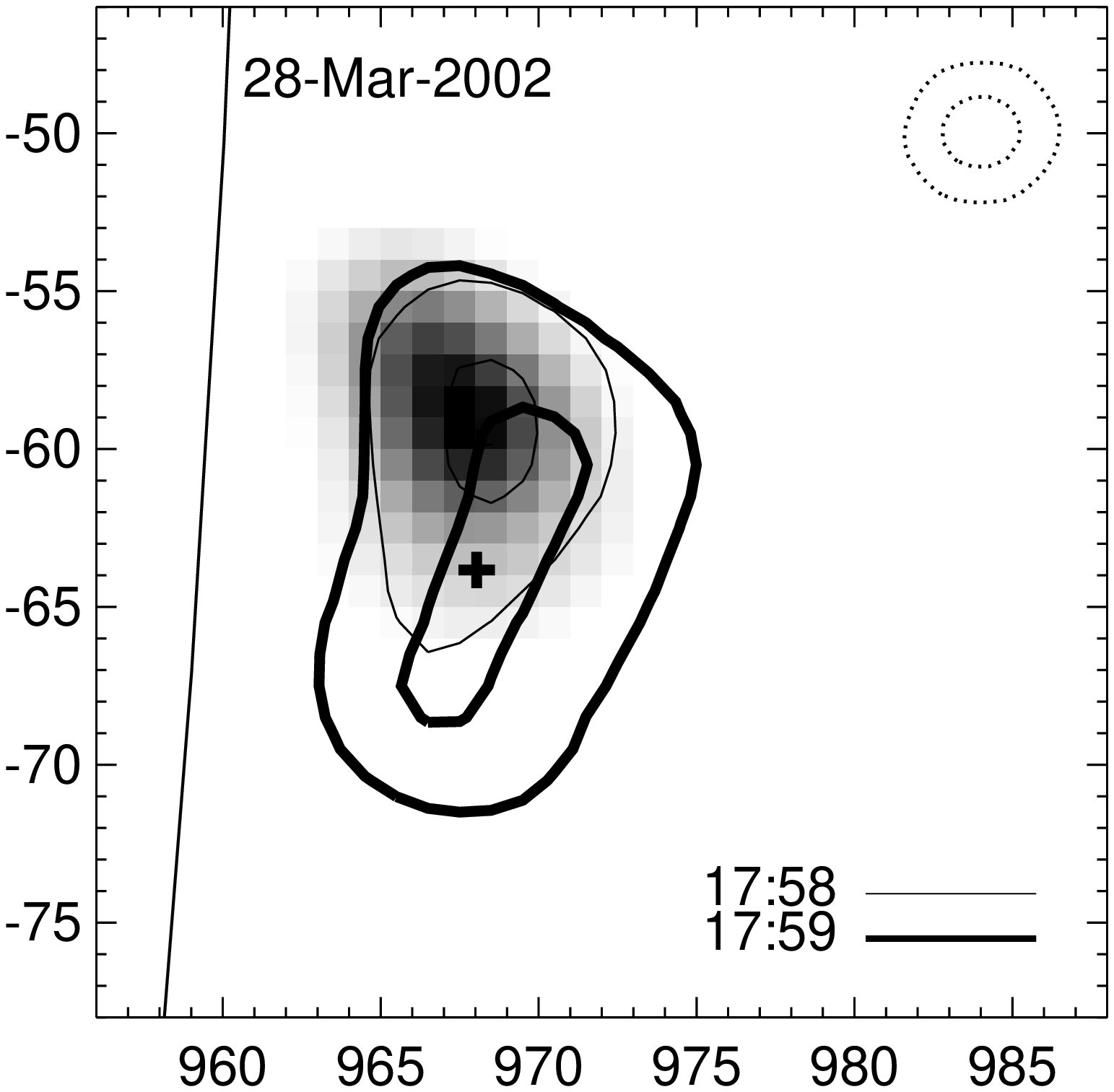}
\plotone{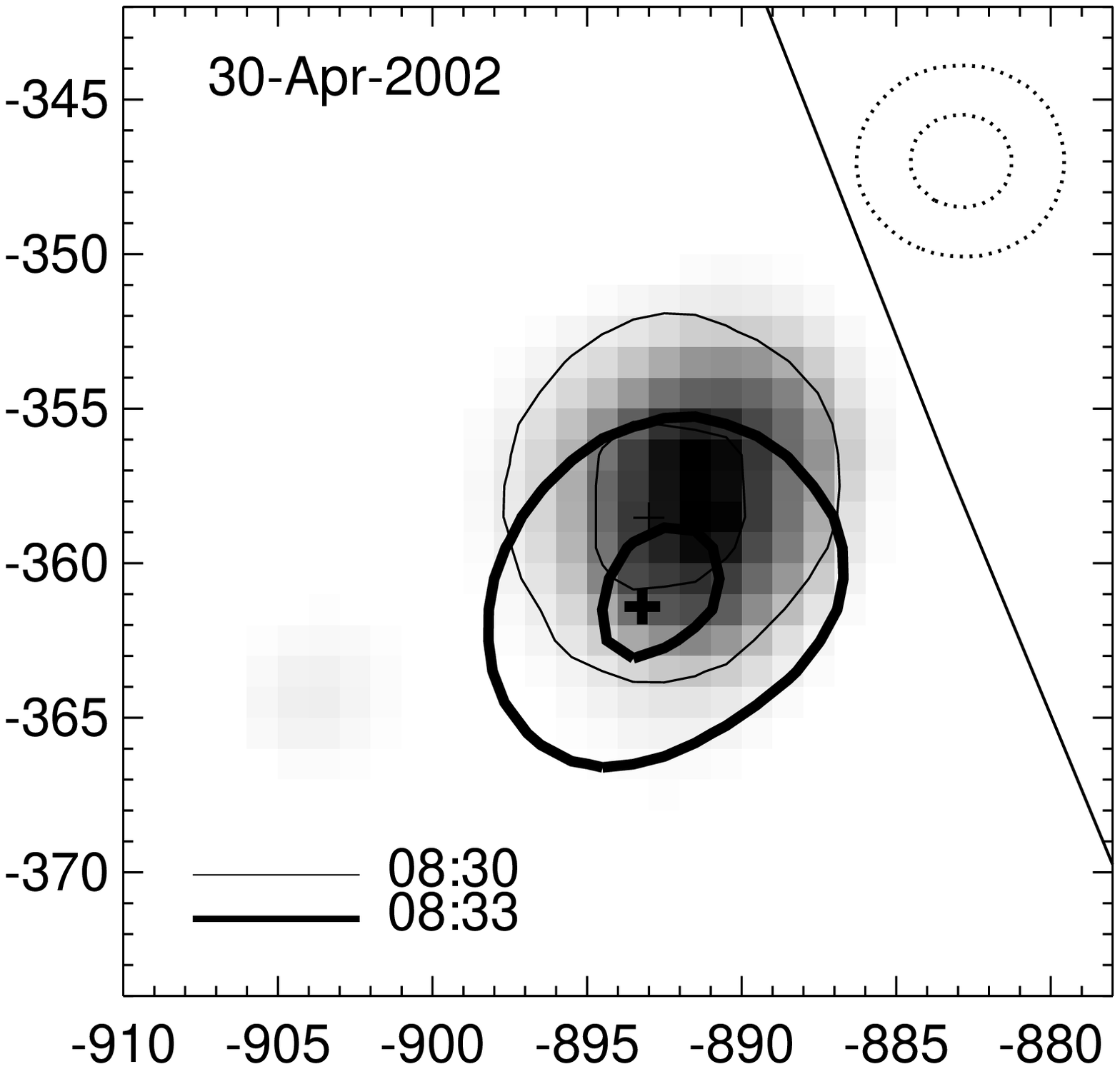}
}
\caption{
Same as Figure \ref{images2.ps} but for the two corona events on March 28 and April 30.
}
\label{images3.ps}
\end{figure}

\begin{figure}[thb]
\epsscale{1}
\centerline{\plottwo{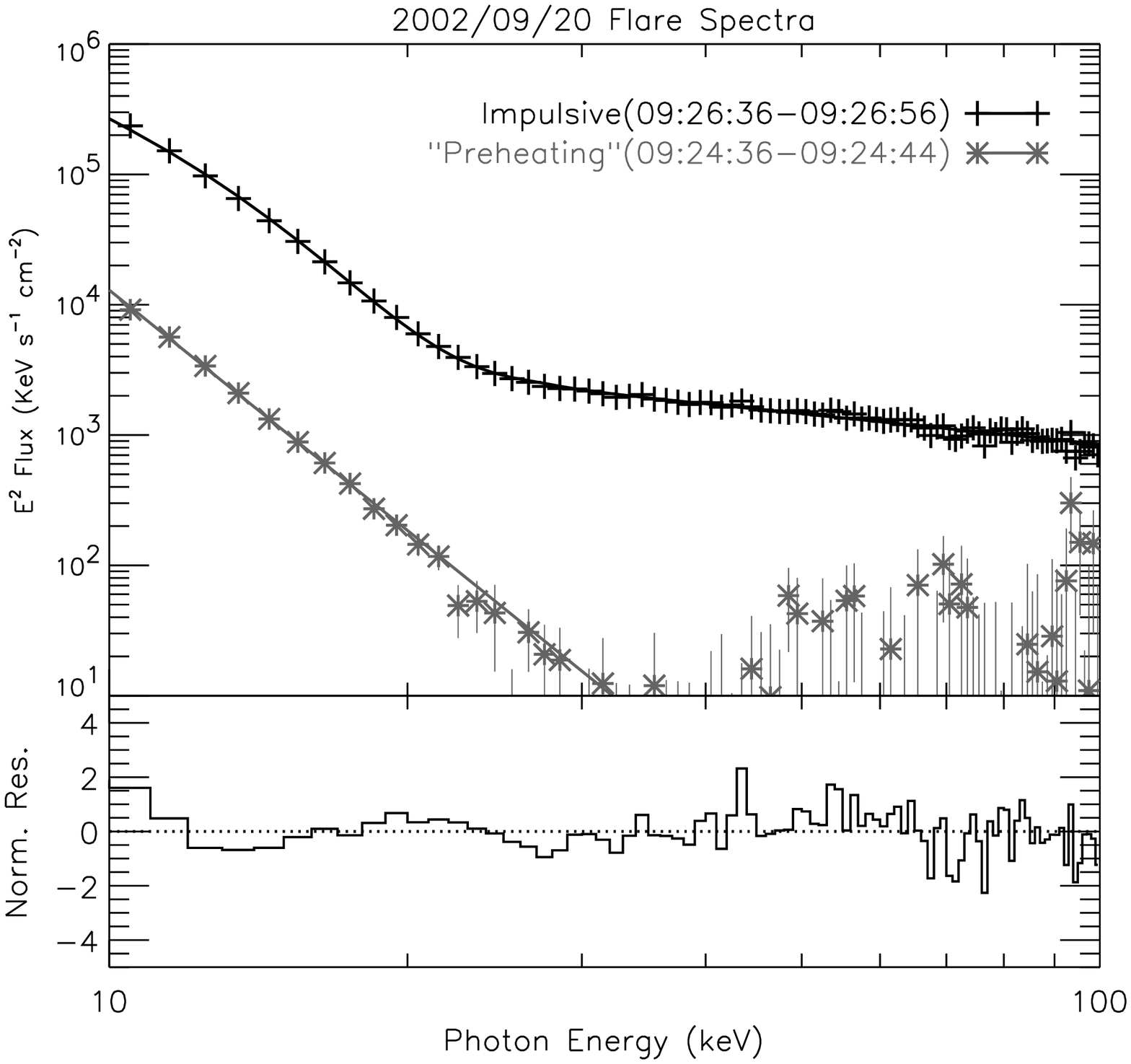}{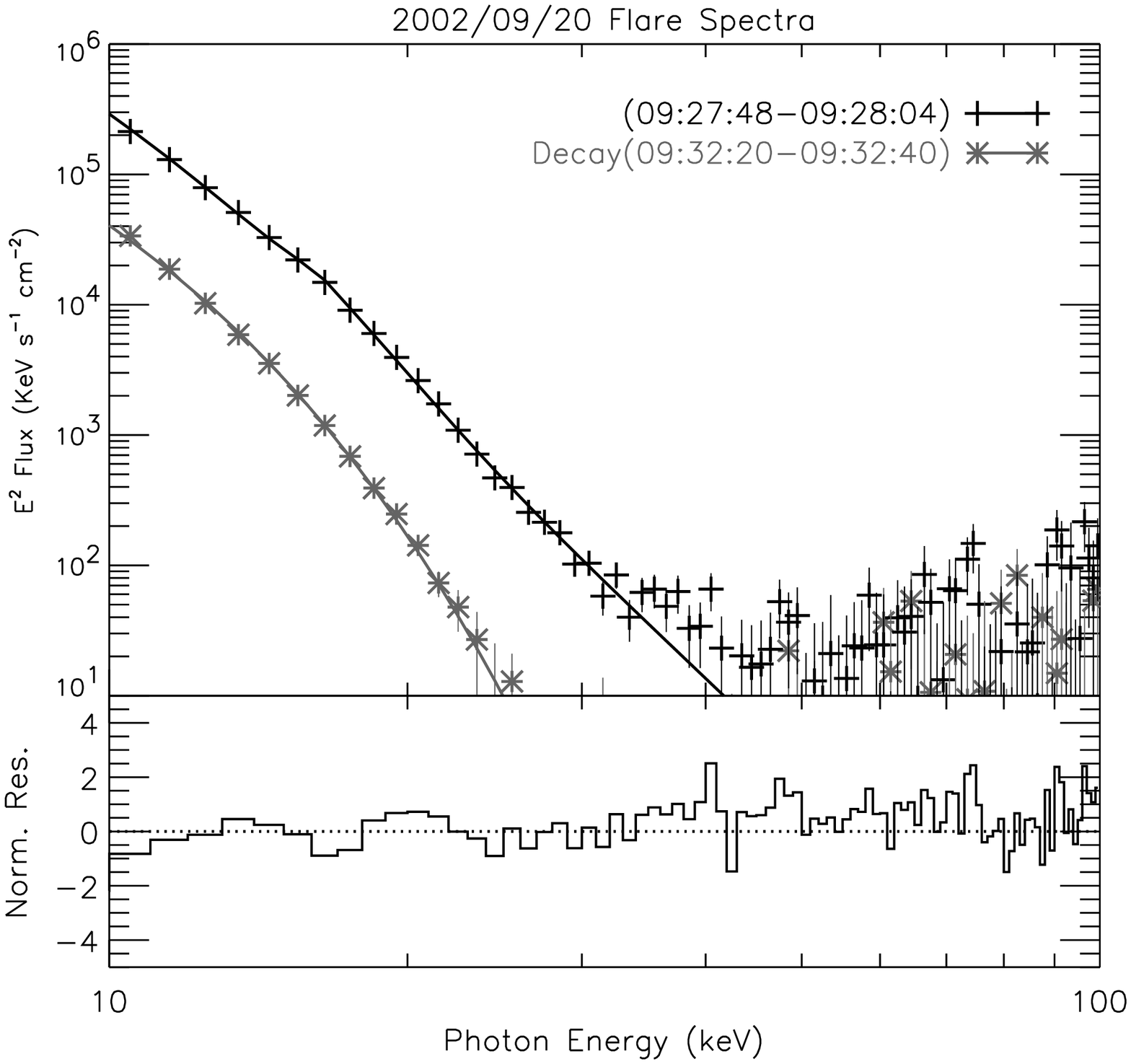}}
\caption{
The {\it RHESSI} total spectrum of the limb flare of September 20 2002. The left panel is for 
the rising phase, fitted with a single power-law (lower), and for the HXR peak (upper), 
fitted with a thermal plus a power-law model, whose residual in units of the standard deviation is 
shown in the lower panel.  The right panel is for the early decay phase (upper), fitted with a 
thermal plus a power-law model with the residuals in the lower panel, and for the late decay phase 
(lower), fitted with a pure thermal model. The corresponding time intervals are indicated in the 
figure and in the light curves by the arrows in the left panel of Figure \ref{cooling.ps}. 
}
\label{specs.ps}
\end{figure}

\begin{figure}[thb]
\epsscale{1}
\centerline{\plottwo{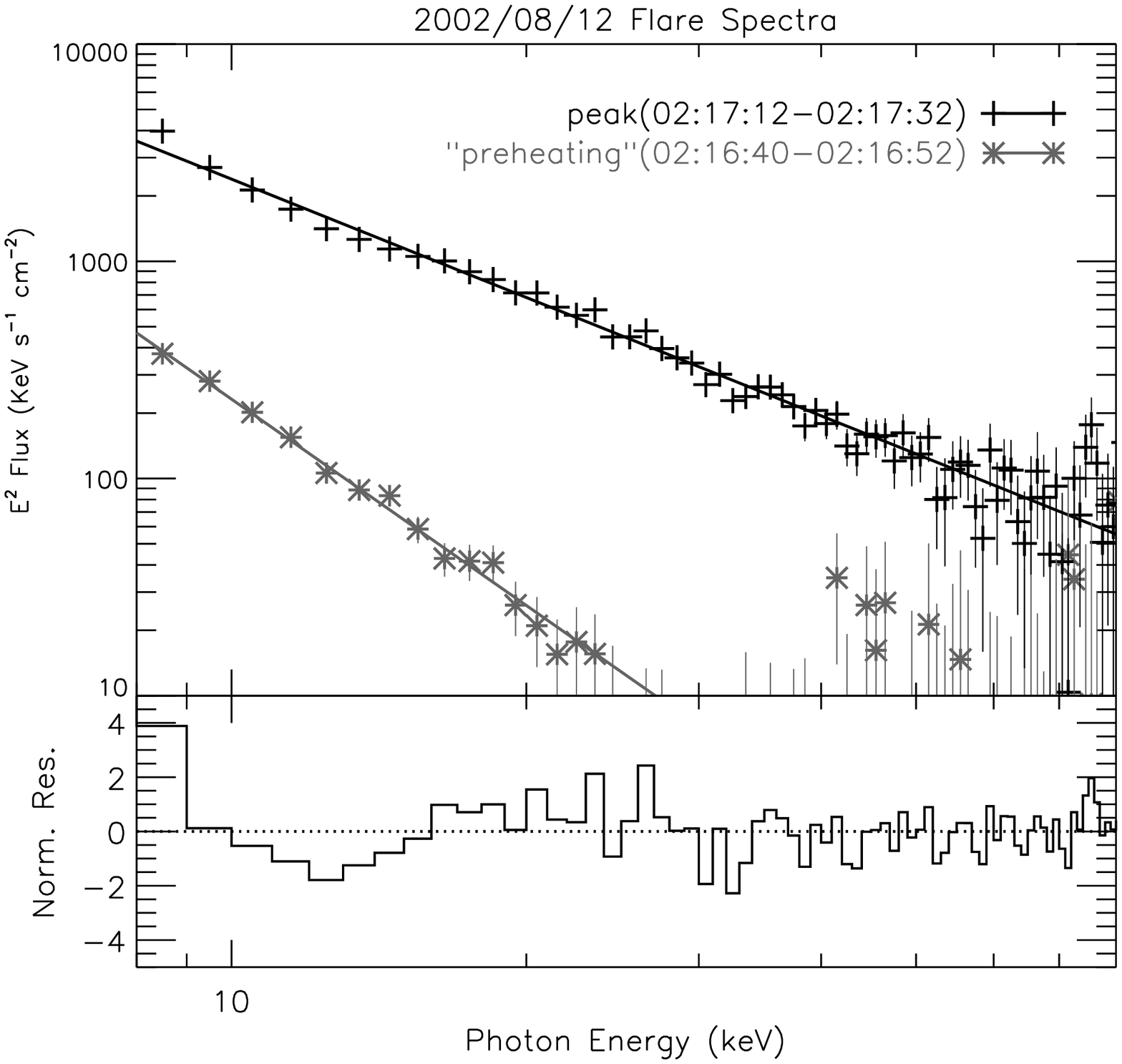}{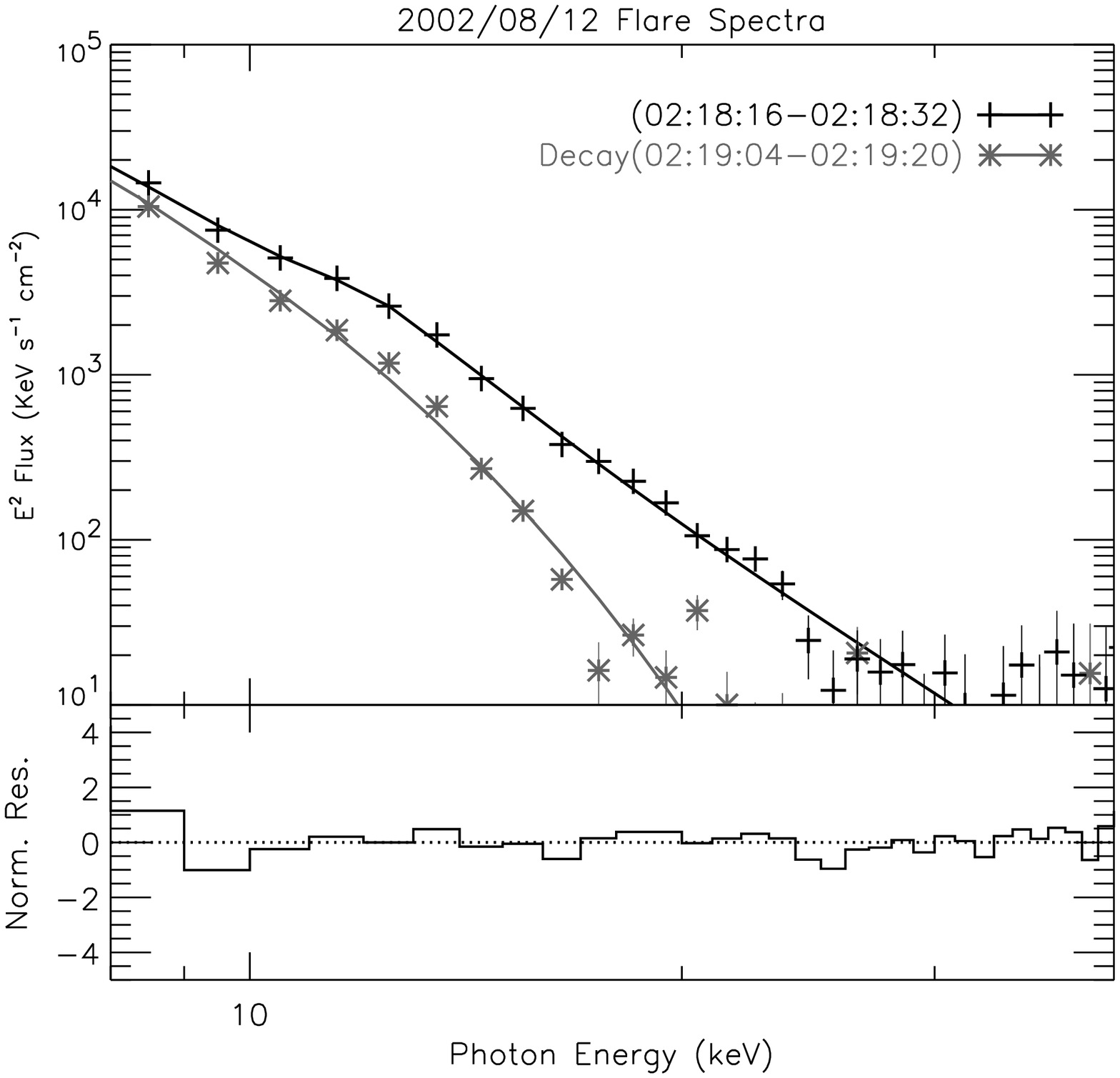}}
\caption{Same as Figure \ref{specs.ps} but for the August 12 flare. Here the peak time spectrum 
is also fitted with a single power-law model. The corresponding 
light curves are shown in right panel of Figure \ref{cooling.ps}.
}
\label{specs2.ps}
\end{figure}

\begin{figure}[thb]
\epsscale{1}
\centerline{\plottwo{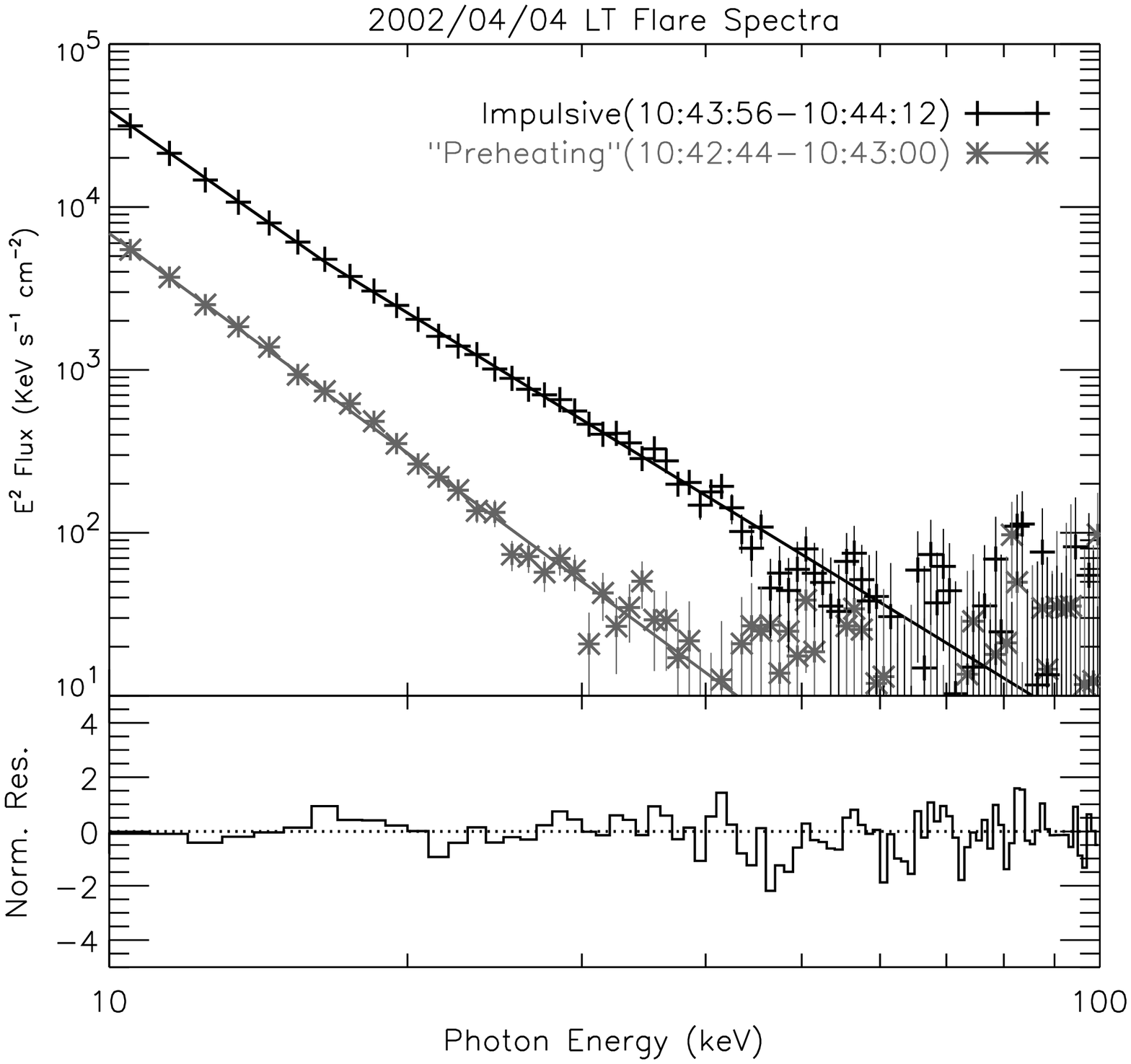}{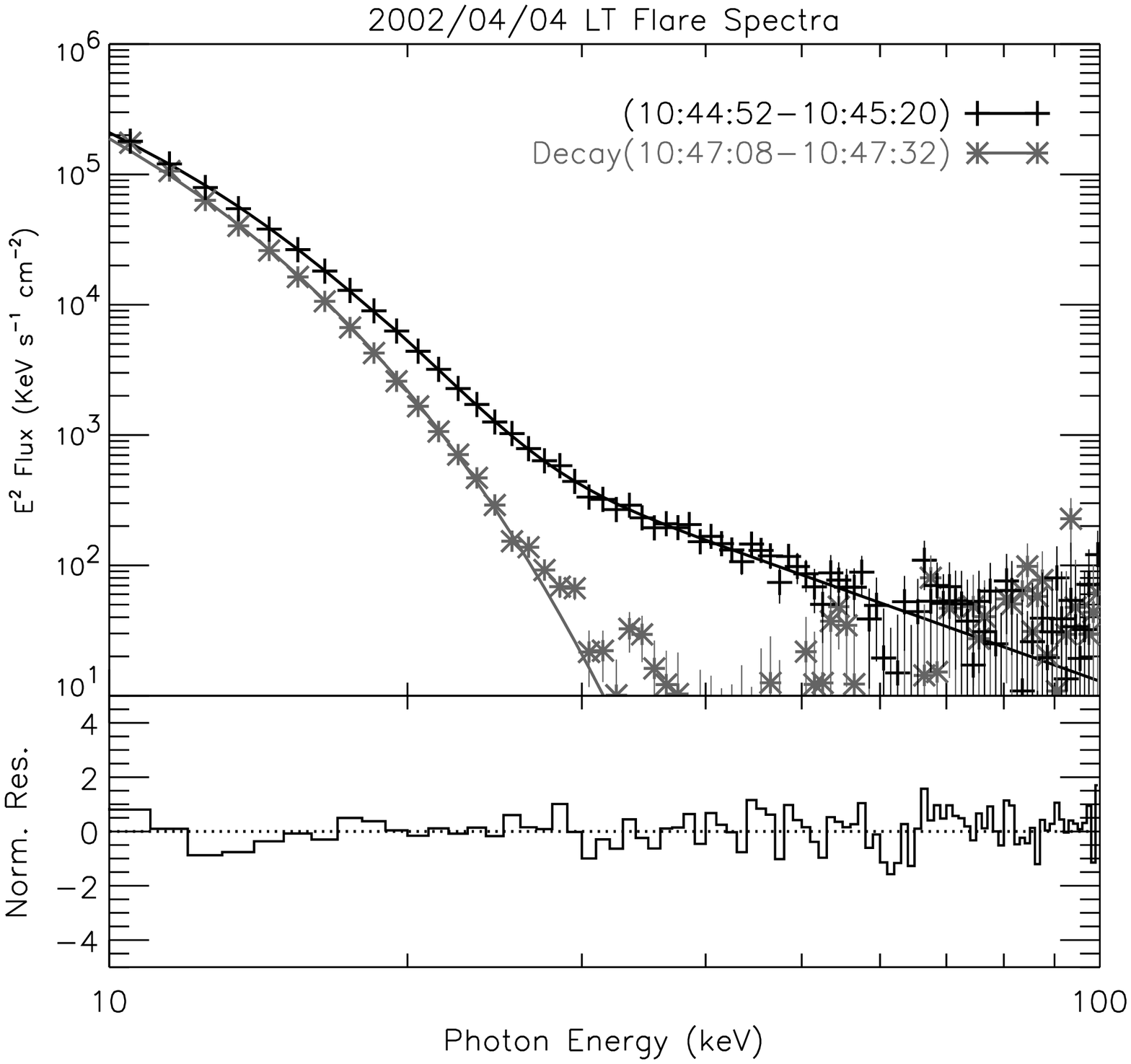}}
\caption{
Same as Figure \ref{specs2.ps} but for the corona event b on April 4 2002. The corresponding 
light curves are shown in right panel of Figure \ref{cooling2.ps}.
}
\label{specs3.ps}
\end{figure}

\begin{figure}[thb]
\epsscale{1}
\centerline{
\plottwo{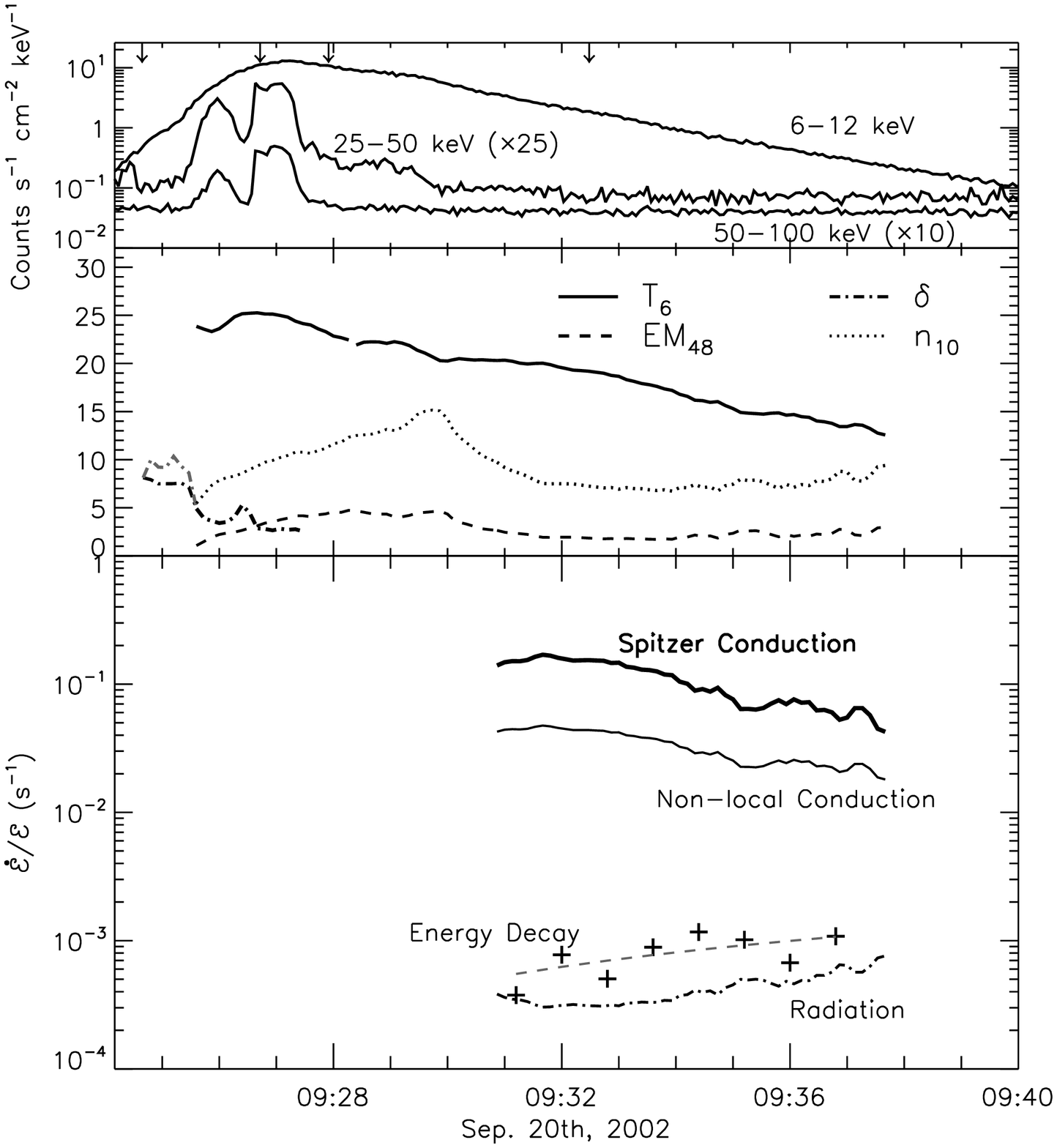}{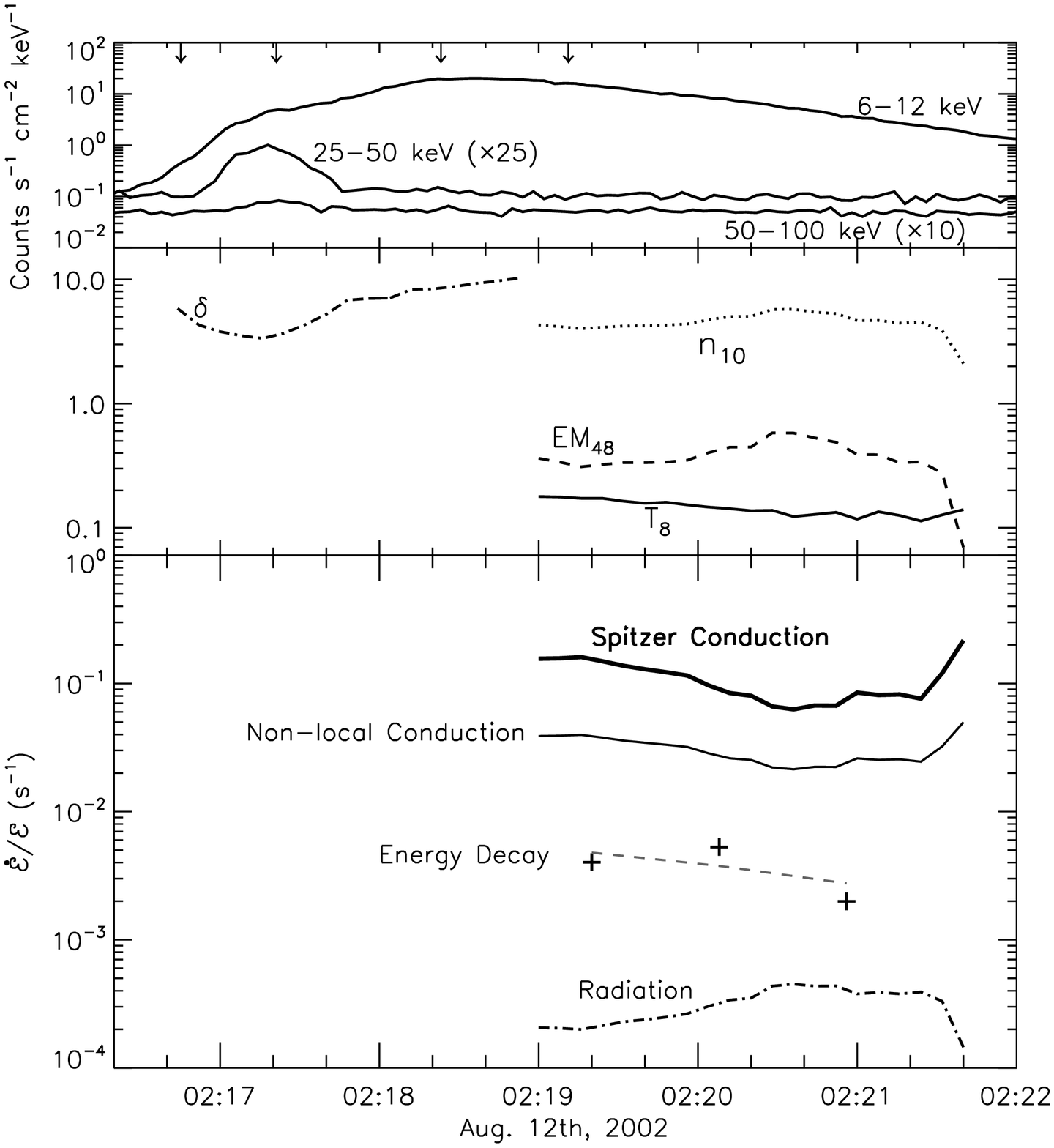}
}
\caption{
 {\it RHESSI} light curve and the evolution of model parameters of the limb flares of September 20 
2002 (left panel) and August 12 2002 (right panel). The top panels give the 6-12, 25-50, and 
50-100 keV light curves with the higher energy ones shifted by the factors indicated in the 
figures for clarity. The arrows on the top indicate the times for the spectral fits shown in 
Figures \ref{specs.ps}-\ref{specs2.ps}. The second panels show the evolution of the emission measure $EM$ 
(dashed line) and temperature $T$ (solid line) for the thermal component, and the corresponding gas 
density $n$ (dotted line) read from the observed source size. Here all quantities are given in cgs 
units and are scaled by the order of magnitude indicated in the subscript. The spectral indexes for either 
a single power-law fit or a broken 
power-law fit are also indicated with the dash-doted lines (gray for higher energy component and black 
for lower one for the broken power-law model). The third panel gives the observed energy decay rate 
(pluses) and those due to the radiative cooling (dot-dashed), Spitzer conduction (thick solid) and 
non-local conduction (thin solid). 
}
\label{cooling.ps}
\end{figure}

\begin{figure}[thb]
\epsscale{1}
\centerline{\plottwo{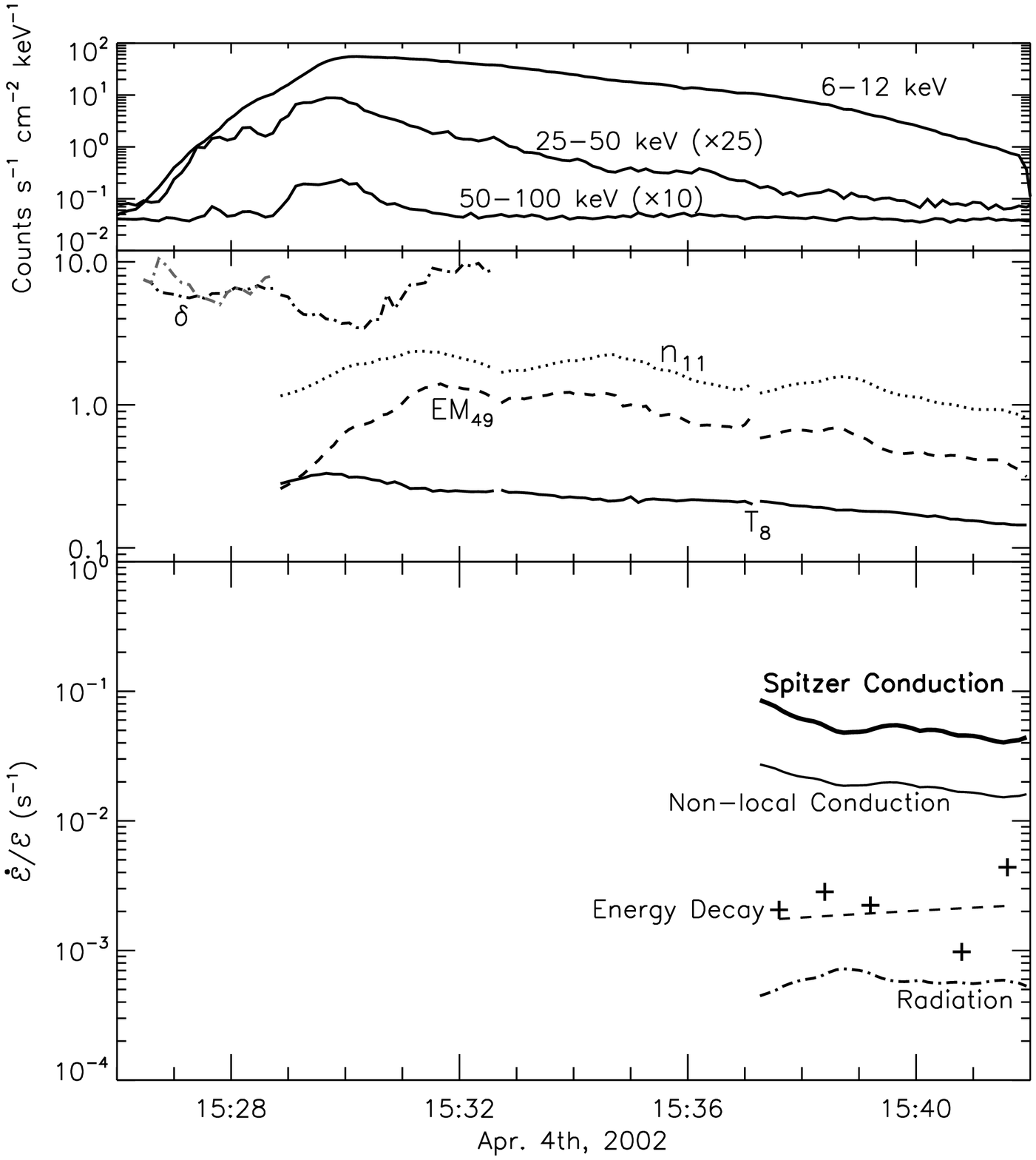}{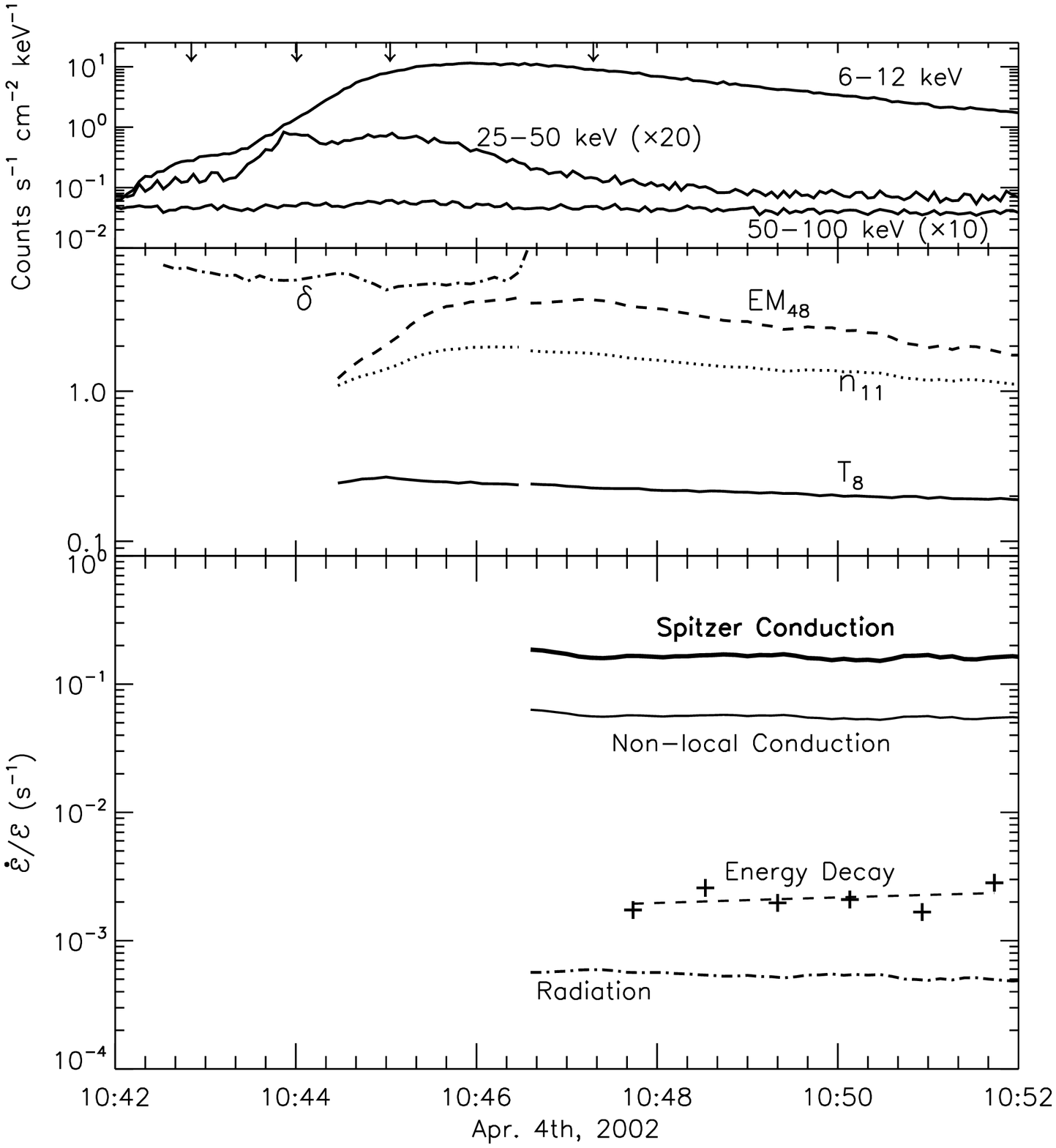}}
\caption{
Same as Figure \ref{cooling.ps} but for the partially occulted flare a ({\it 
Left}) and the occulted flare b ({\it Right}) on April 4 2002.}
\label{cooling2.ps}
\end{figure}

\begin{figure}[thb]
\epsscale{1}
\centerline{\plottwo{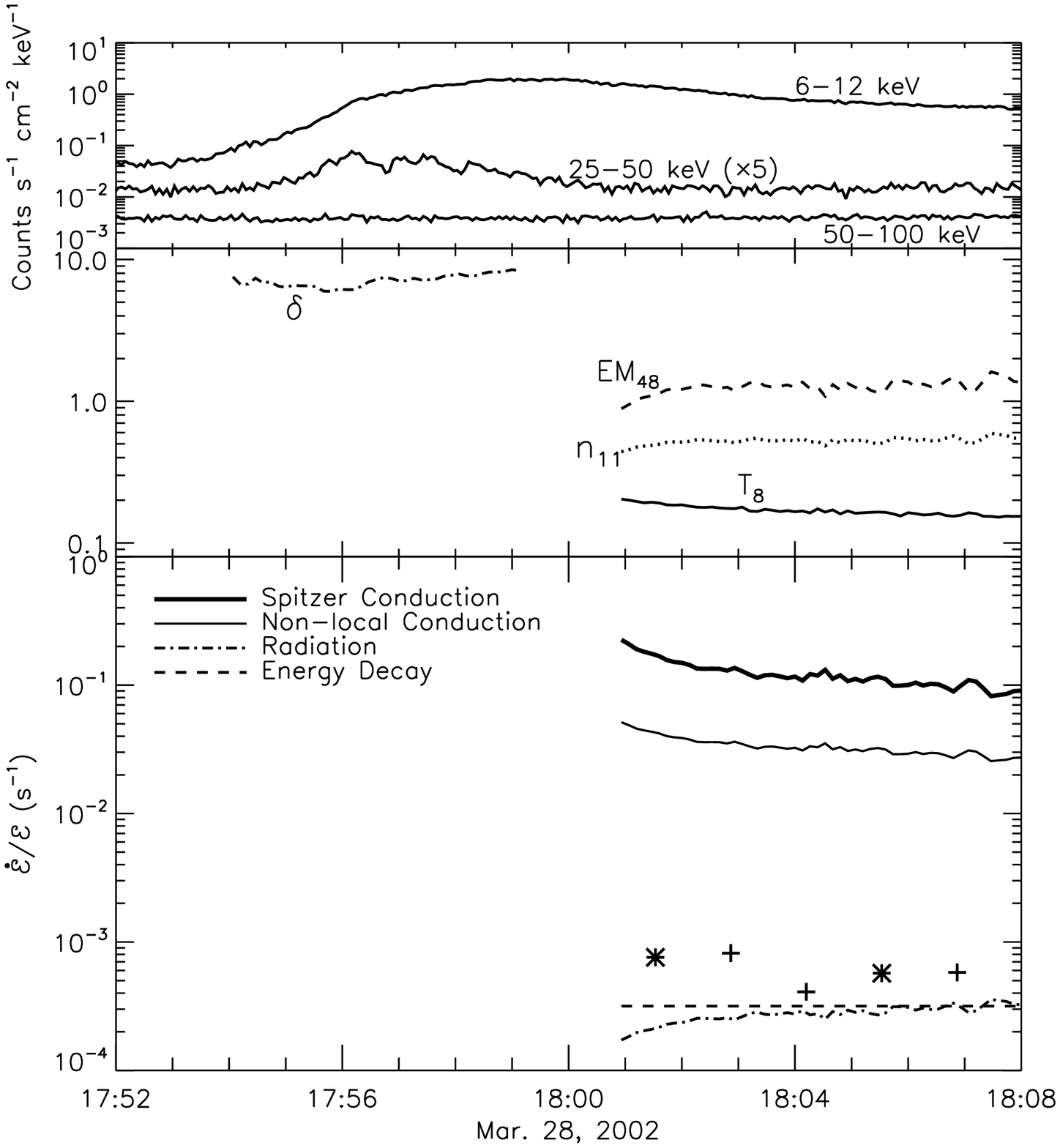}{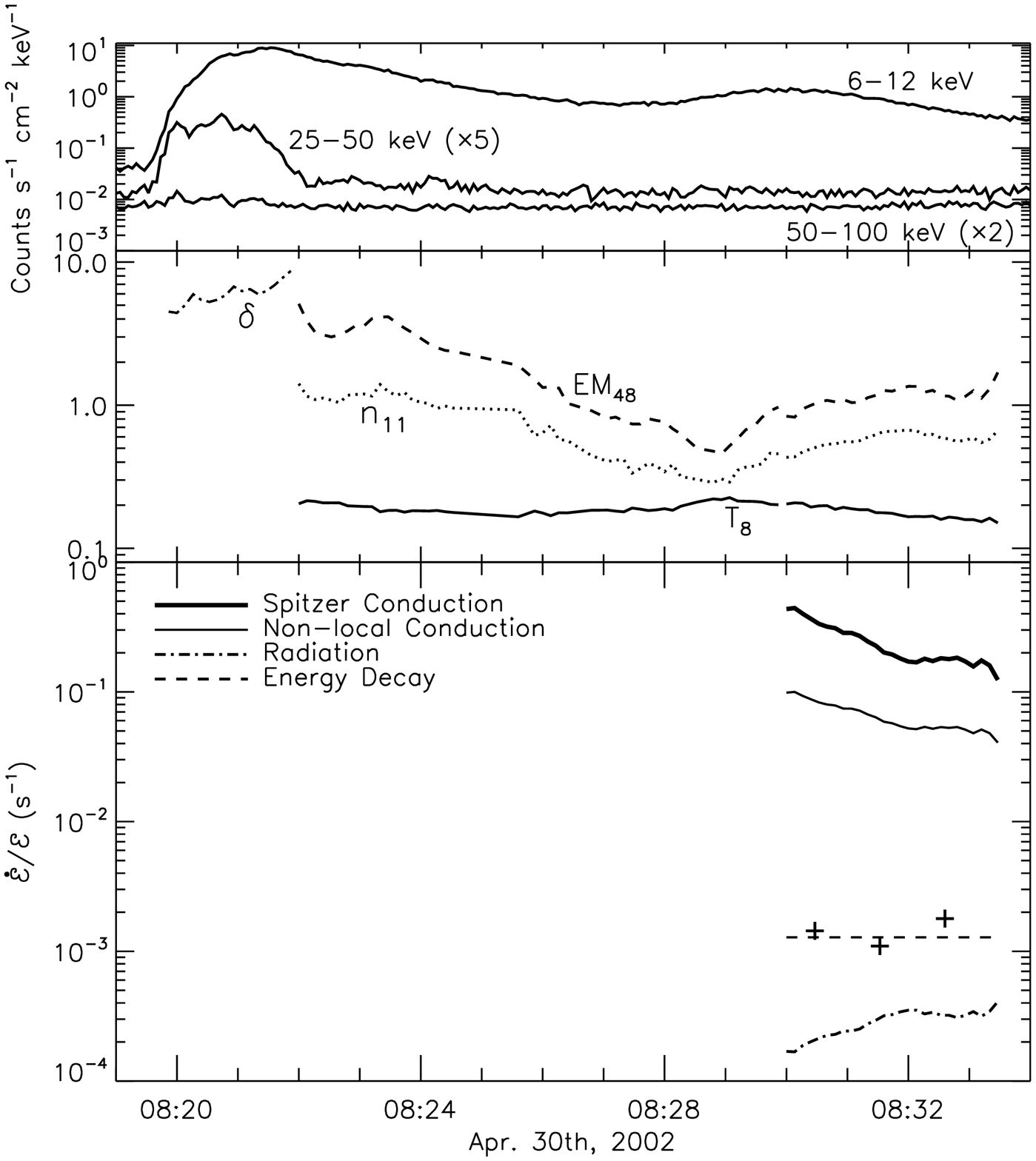}}
\caption{
Same as Figure \ref{cooling.ps} but for the flare on March 28 2002 ({\it Left}) and the
flare on April 30 2002 ({\it Right}). The star signs indicate the energy increase rate during the 
periods.
}
\label{cooling3.ps}
\end{figure}

We have carried out imaging and spectroscopy studies of the evolution of 6 simple limb flares, 
some of which  reveal several interesting features of the LT source as enumerated below: 
\begin{itemize}
\item A distinct coronal LT source (Figures \ref{images1.ps}-\ref{images3.ps}) is observed for 
each flare. The source appears upon the onset of the flare and exists until the flare dies out. It 
is relatively stable in both shape and size and, in some cases, moves systematically toward high 
latitudes in the decay phase.
\item For a few flares with their FPs completely occulted by the solar limb, the spectrum of 
the LT source can be studied without invoking imaging spectroscopy. Above 10 keV, the spectrum is  
dominated by a steep power-law during the preheating and impulsive phases (Fig. \ref{specs.ps}). 
A thermal component appears after the impulsive HXR peak and becomes more and more prominent as 
the flare proceeds (Fig. \ref{cooling.ps}).
\item The LT source spectrum can be fitted with a thermal model in the late decay phase. The 
cooling rate of the LT source is generally higher than the radiative cooling rate but much 
lower than the conductive cooling rate predicted by Spitzer or the so-called non-local 
conductivity, requiring a suppressed conductivity and/or continuous heating.
\item Imaging of this LT source indicates that the thermal hot plasma is confined within a small 
region near the LT and does not extend to the FP regions. 
\end{itemize}

We have considered several possibilities to explain these observations and carried out a detailed 
investigation of the energy balance in the post impulsive loop:
\begin{itemize}
\item A small filling factor can not explain the discrepancy between the high decay rate expected 
theoretically and the observed cooling rate, when the effects of both conduction and 
radiation are included. It also can not produce the confinement of the LT source.
\item For loop  models with a uniform conductivity because the conductive heat flux and the 
radiative cooling rate are very sensitive to the temperature, the loop must have a nearly uniform 
temperature profile to balance efficient radiative cooling near the cold FPs. This results in LT 
sources extending much closer to the FPs than those observed by {\it RHESSI}.
\item The suppression of conduction therefore must be confined to the LT region. Moreover for some 
flares (i.e. the September 20 flare) continuous energy input is also required to balance the 
rapid cooling in the loop legs. For flares with relative longer decay time the energy input in the 
decay phase can exceed that in the impulsive phase, and most of these energy are radiated away 
near the FPs through the optical and UV emission lines. This prediction can be tested with 
broad band observations. The fluctuations in the energy decay rate and occasionally observed 
energy increase of some flares also suggest a continuous heating process. 
\item We suggest that plasma waves or turbulence may account for both these aspects. The 
scattering of high energy electrons by waves can reduce the conduction. The particles can also gain 
energy from the waves, which would account for the heating.
\item Assuming the scattering time scale independent of the particle velocity and a resonant 
wave-particle coupling, which gives a heating rate inverse proportional to the square of the 
particle velocity, we showed that observations can be used to determine the size of the turbulence 
region and the degree of conduction suppression and plasma heating.
\end{itemize}

This is a new result indicating that plasma waves or turbulence, which appears to play important 
roles in the acceleration of particles during the impulsive phase (Miller et al. 1987; 
Petrosian \& Liu 2004), be generated continuously even during the decay phase, presumably at a 
lower level. We note that according to Petrosian \& Liu (2004) a low level of turbulence means 
a slow plasma heating with little particles accelerated to high energies. The high gas density and 
temperature in the gradual phase also make the plasma heating more dominant. More detailed studies 
of wave-particle interactions, i.e. the scattering and acceleration of particles by PWT, will help 
us better understand the relevant phenomena, uncovering the energization mechanism of solar flares 
eventually.

{\bf Acknowledgments} 

The work is supported by NASA grants NAG5-12111, NAG5 11918-1, and NSF grant ATM-0312344. We
are indebted to S. Krucker and the {\it RHESSI} team for numerous help. We also would like to 
thank T. Metcalf, G. Hurford for helpful discussions and suggestions, and are grateful to K. 
Tolbert, R. Schwartz, and J. McTiernan for their help on the data analyses.

{}

\clearpage

\vspace {0.1in}
\begin{center}
\begin{table}[ht]
\caption{
The ratios of the energy decay rates expected from the radiative, Spitzer conductive, and 
non-local conductive cooling to the observed one. The radiative ones are a bit smaller than 
unity while the conductive ones are much larger than one. The minimum energy decay rates 
due to the combined cooling of radiation and Spitzer conduction are obtained by adjusting the 
volume filling factor $\eta$. The corresponding critical suppression factor $\eta_{\rm cr}$ is 
given in 
the last column. The fact that this minimum values are always larger than one shows that 
variations in $\eta$ can 
not explain the observations.
}
\vspace {0.05in}
\begin{tabular}{lccccc}
\tableline\tableline
{\it RHESSI} & \multicolumn{4}{c} {Ratio of Model Predicted to Observed Temperature 
Decay Rate}& Filling Factor\\ \cline{2-5} 

Flare \# & Radiation & Spitzer& Non-local & Radiation \& Conduction &\\
	&	& Conduction & Conduction & Minimum & $\eta_{\rm cr}(10^{-3})$ \\
\tableline
2032819 &$ 0.87 $&$ 386 $&$ 105 $&$ 37 $& 2.3\\
2040411 &$ 0.28 $&$ 87 $&$ 30 $&$ 9.9 $& 3.3 \\
2040413 &$ 0.38 $&$ 34 $&$ 12 $&$ 7.2 $& 11\\
2043004 &$ 0.22 $&$ 189 $&$ 51 $&$ 13 $& 1.2\\
2081203 &$ 0.084 $&$ 29 $&$ 8.0 $&$ 3.1 $& 2.9\\
2092002 &$ 0.52 $&$ 130 $&$ 41 $&$ 17 $& 4.0\\

\tableline
\end{tabular}
\end{table}
\end{center}

\clearpage

\end{document}